\documentstyle[12pt]{article}

\begin{document}

\baselineskip=19pt plus 0.2pt minus 0.1pt

%%%%%%%%%%% Private Macros %%%%%%%%%%%%%
\makeatletter

\def\@bgnmark{<}
\def\@endmark{>}
\def\WKht{.85}
\def\WKsep{.4}
\def\WKrule{.03}
\newcount\@bgncnt
\newcount\@endcnt
\newcount\@h@ight
\newcount\TempCount
\newif\if@Exist
%\newdimen\@bgnpt
%\newdimen\@endpt
\newdimen\@tempdimc
\newdimen\@tempdimd
\newdimen\h@ight % unit height of wicksymbol in unit of ex
\newdimen\w@dth  % thickness of the rule of wicksymbol
\def\sqrt{\radical"270370}%\def\sqrt{\@@sqrt}
%\def\first#1{\expandafter\@car#1\@nil}
%\def\rest#1{\expandafter\@cdr#1\@nil}

%% Some functions to cut a part of strings out of
%% a string = MAE {< or >} MARK FRONT USIRO
\def\SEPbgn#1<#2#3#4\@@{\xdef\@MAE{#1}\xdef\@MARK{#2}
\xdef\@FRONT{#3}\xdef\@USIRO{#4}}
\def\SEPend#1>#2#3#4\@@{\xdef\@MAE{#1}\xdef\@MARK{#2}
\xdef\@FRONT{#3}\xdef\@USIRO{#4}}
\def\c@lc{% \@tempdima: position, \@tempdimb: letter-height
 \setbox0=\hbox{$\displaystyle \@FRONT$}
 \@tempdima\wd0 \@tempdimb\ht0
 \settowidth{\@tempdimc}{$\displaystyle \@MAE$}
 \settowidth{\@tempdimd}{$\displaystyle \@list$}
 \divide\@tempdima by2 \advance\@tempdima by \@tempdimc
 \advance\@tempdima by \@tempdimd}

%%%%%%%%%% A useful macro \@dblfor for general use %%%%%%%%%%%%%%%%
%% USAGE:
%%    \@dblfor#1;#2:=#3\do#4
%%
%% For the `pair-ed' data of the form #3:(A1,B1, ...,An,Bn, ..., AN,BN,)
%%                     Note that a comma exists at the end of list ---^
%% You can refer to An by name #1, to Bn by name #2
%% and do the routine #4 for n=1,2,3,...,N successively.
%%------- Essentially a variation of the LaTeX macro \@tfor -------
\def\@dblfornoop#1\@@#2#3#4{}
\def\@dblfor#1;#2:=#3\do#4{\xdef\@fortmp{#3}\ifx\@fortmp\@empty \else%
 \expandafter\@dblforloop#3\@nil,\@nil,\@nil\@@#1#2{#4}\fi}
\def\@dblforloop#1,#2,#3\@@#4#5#6{\def#4{#1} \def#5{#2}%
 \ifx #4\@nnil \let\@nextwhile=\@dblfornoop \else%
 #6\relax \let\@nextwhile=\@dblforloop\fi\@nextwhile#3\@@#4#5{#6}}
%%%%%%%%%%%%%%%%%%%%%%%%%%%%%%%%%%%%%%%%%%%%%%%%%%%%%%%%%%%%%%%%%%%

%%%%%%%% next uses this macro \@dblfor %%%%%%%%%%%
\def\fin@endpt#1#2{% find the endpoint with label #1 from the list #2
%                  % #2=(label1,endpoint1,label2,endpoint2, ..... ,)
\@dblfor\MemBer;\NextmemBer:=#2\do{\def\@bject{#1}%
 \if \MemBer\@bject \xdef\@endpt{\NextmemBer} \@Existtrue\fi}}%
%%%%%%%% next uses the standard LaTeX macro \@tfor %%%%%%%%%%%
\def\fin@h@ight#1#2{% find the #1-th height h#1 from the list #2
%                   % #2=(h1 h2 ...) with no commas between hn's.
 \@tempcnta\z@%
 \@tfor\MEmber:=#2\do{\advance\@tempcnta\@ne%
 \ifnum \@tempcnta=#1 \@h@ight=\MEmber\fi}}

%%%%% definition of wicksymbol %%%%%%%%%%%%%%%%%%%%
\def\wicksymbol#1#2#3#4#5{% #1(pos),#2(ht):bgnpnt
%                         % #3(pos),#4(ht):endpnt  #5:height
% Absolute height version
 \@tempdima=#3 \advance\@tempdima-#1%
 \@tempdimc=#5\h@ight \@tempdimb=\@tempdimc \advance\@tempdimb-\w@dth%
 \@tempdimd=#2 \advance\@tempdimd-1.587ex
 \hskip#1%
 \vrule height \@tempdimc width\w@dth depth-\@tempdimd \kern-\w@dth%
 \vrule height \@tempdimc width\@tempdima depth-\@tempdimb\kern-\w@dth%
 \@tempdimd=#4 \advance\@tempdimd-1.587ex
 \vrule height \@tempdimc width\w@dth depth-\@tempdimd}

\def\first#1{\expandafter\@mae#1\@nil}
\def\secnd#1{\expandafter\@ato#1\@nil}
\def\@mae#1;#2\@nil{#1}
\def\@ato#1;#2\@nil{#2}

%%%%% Definition of \wick %%%%%%%%%%%%%
\def\wick#1#2{%
 \h@ight=\WKht ex \w@dth=\WKrule em%
 \def\@wickdata{} \def\bgnend@list{} \@bgncnt\z@ \@endcnt\z@%
 \def\@list{} \def\bgnp@sition{} \def\endp@sition{}%
%% Make \bgnp@sition, \endp@sition and \@list
 \xdef\str@ng{#2}
 \@tfor\m@mber:=#2\do{%
 \ifx\m@mber\@bgnmark \advance\@bgncnt\@ne
  \expandafter\SEPbgn\str@ng\empty\@@ \c@lc
  \xdef\bgnp@sition{\bgnp@sition\@MARK,\the\@tempdima;\the\@tempdimb,}
  \xdef\@list{\@list\@MAE\@FRONT}
  \xdef\str@ng{\@USIRO}\fi
 \ifx \m@mber\@endmark \advance\@endcnt\@ne
  \expandafter\SEPend\str@ng\empty\@@ \c@lc
  \xdef\endp@sition{\endp@sition\@MARK,\the\@tempdima;\the\@tempdimb,}
  \xdef\@list{\@list\@MAE\@FRONT}
  \xdef\str@ng{\@USIRO}\fi}
  \xdef\@list{\@list\@USIRO}
%% Warning an error
 \ifnum\@bgncnt=\@endcnt \else%
 \@latexerr{The numbers of `<' and `>' do not match}%
 {You have written different numbers of < and >}\fi%
%% Count the number of height-numbers
 \TempCount\z@ \@tfor\mmbr:=#1\do{\advance\TempCount\@ne}%
 \ifnum\@bgncnt=\TempCount \else%
 \@latexerr{The number of numbers in the first argument is different
 with that of contractions <...>}%
 {Give the same numbers of heights as the contractions <...>}\fi%%
%% finally write wick-symbols
 \mathop{\vbox{\m@th\ialign{##\crcr\noalign{\kern\WKsep ex}%
 $\m@th \TempCount\z@%
 \@dblfor\member;\nextmember:=\bgnp@sition\do{% make \@wickdata
 \advance\TempCount\@ne \xdef\@bgnpt{\nextmember}%
 \@Existfalse%
 \fin@endpt{\member}{\endp@sition}%
 \if@Exist \else \@latexerr{The begin-mark `<\member' has no
corresponding end-mark `>\member'}{You should write coinciding label
like <\member .. >\member}\fi%
 \fin@h@ight{\TempCount}{#1}%
 \setbox0=\hbox{%
 $\wicksymbol{\first\@bgnpt}{\secnd\@bgnpt}{\first\@endpt}%
 {\secnd\@endpt}{\@h@ight}$\hss}
 \dp0\z@ \wd0\z@ \box0%
% \hbox to0pt{$\wicksymbol{\@bgnpt}{\@endpt}{\@h@ight}$\hss}
 }$\crcr\noalign{\kern\WKsep ex\nointerlineskip}%
 \setbox0=\hbox{$\displaystyle\@list$}\ht0=1.587ex%
 \box0\crcr}}}\limits}
%%%%%%%%%%%% End of wick.sty %%%%%%%%%%%%%%%%%%%%%%%%%%%%%%%%%%

\@addtoreset{equation}{section}
\renewcommand{\theequation}{\thesection.\arabic{equation}}
\renewcommand{\thefootnote}{\fnsymbol{footnote}}
\newcommand{\nn}{\nonumber}
\renewcommand{\star}{*}
\newcommand{\tr}{\mathop{\rm tr}}
\newcommand{\QB}{Q_{\rm B}}
\newcommand{\wtQB}{\widetilde{Q}_{\rm B}}
\newcommand{\Half}{\frac{1}{2}}
\newcommand{\bra}[1]{\left\langle #1\right|}
\newcommand{\ket}[1]{\left| #1\right\rangle}
\newcommand{\VEV}[1]{\left\langle #1\right\rangle}
\newcommand{\braket}[2]{\VEV{#1 | #2}}
\newcommand{\dL}{\delta_\Lambda}
\newcommand{\Sz}{S_0}
\newcommand{\Ss}{S_{\rm source}}
\newcommand{\abs}[1]{\left\vert #1\right\vert}
\newcommand{\D}{W}
\newcommand{\ac}{\overline{c}}
\renewcommand{\a}{\alpha}
\newcommand{\B}[1]{B_{\rm #1}}
\newcommand{\calO}{{\cal O}}
\newcommand{\calD}{{\cal D}}
\newcommand{\p}{\partial}
\newcommand{\Drv}[2]{\frac{d#1}{d#2}}
\newcommand{\Pdrv}[2]{\frac{\partial #1}{\partial #2}}
\newcommand{\pdrv}[2]{\partial #1/\partial #2}
\newcommand{\zzero}{\zeta(0)}
\newcommand{\F}{\frac{1-F}{1+F}}
\newcommand{\f}{\frac{1}{1+F}}
\newcommand{\tp}[1]{{#1}^{T}}
\newcommand{\dLm}{\delta_{\Lambda_-}}
\newcommand{\dLp}{\delta_{\Lambda_+}}
\newcommand{\dLt}{\delta_{\Lambda_t}}
\newcommand{\dD}{\delta_{\rm D}}
\newcommand{\ie}{{\em i.e.}}
\newcommand{\wt}[1]{\widetilde{#1}}
\newcommand{\isqrt}[1]{\frac{1}{\sqrt{#1}}}
\newcommand{\G}{{\cal G}}
\newcommand{\wtNFP}{\widetilde{N}_{\rm gh}}
\newcommand{\NFP}{N_{\rm gh}}
\newcommand{\ds}{\displaystyle}
\renewcommand{\S}[1]{S_{\rm #1}}
\newcommand{\BOX}{\lower2pt\hbox{\Large$\Box$}}
\newcommand{\e}{\varepsilon}
\newcommand{\N}[2]{\overline{N}_{#1}^{\,#2}}
\newcommand{\bm}[1]{\mbox{\boldmath $#1$}}
\renewcommand{\P}{\bm{P}}
\newcommand{\tlt}{\theta}
\newcommand{\RT}[1]{\frac{e\alpha_{#1}}{\e}}
\newcommand{\IRT}[1]{\frac{\e}{e\alpha_{#1}}}
\newcommand{\sgn}{\mathop{\rm sgn}}
\newcommand{\g}{\gamma}
\newcommand{\ag}{\overline{\gamma}}

\makeatother
%%%%%%%%% End of private macros %%%%%%%%%%%

\begin{titlepage}
\title{
\hfill\parbox{4cm}
{\normalsize KUNS-1438\\HE(TH)~97/05\\{\tt hep-th/9704125}}\\
\vspace{1cm}
D-brane and Gauge Invariance\\ in Closed String Field Theory}
\author{
Koji {\sc Hashimoto}\thanks{{\tt hasshan@gauge.scphys.kyoto-u.ac.jp}}
{}\thanks{
Supported in part by Grant-in-Aid for Scientific
Research from Ministry of Education, Science and Culture
(\#3160).}
{} and
Hiroyuki {\sc Hata}\thanks{{\tt hata@gauge.scphys.kyoto-u.ac.jp}}
{}\thanks{
Supported in part by Grant-in-Aid for Scientific
Research from Ministry of Education, Science and Culture
(\#07640394).}
\\[7pt]
{\it Department of Physics, Kyoto University, Kyoto 606-01}
}
\date{\normalsize April, 1997}
\maketitle
\thispagestyle{empty}

\begin{abstract}
\normalsize

We construct a system of bosonic closed string field theory coupled to
a D-brane. The interaction between the D-brane and closed string field
is introduced using the boundary state which is a function of constant
field strength $F_{\mu\nu}$ and the tilt $\tlt_\mu^i$ of the
D-brane. We find that the gauge invariance requirement on the system
determines the $(F_{\mu\nu},\tlt_\mu^i)$-dependence of the
normalization factor of the boundary state as well as the form of the
purely $(F_{\mu\nu},\tlt_\mu^i)$ term of the action.
Correspondence between the action in the present formalism and the low
energy effective action (bulk + D-brane actions) in the $\sigma$-model
approach is studied.
\end{abstract}

\end{titlepage}
%%%%%%%%%%%%%%%%%%%%%%%%%%%%%%%%%%%%%%%%%%%%%%%%%%%%%%%%%%%%%%%%%%%

\section{Introduction}
\label{sec:intro}

D-branes \cite{DLP,Polchinski,PolchRev}, ``solitonic'' objects
carrying Ramond-Ramond charges, play crucial roles in non-perturbative
understanding of string theories based on dualities \cite{Duality}.
Their existence was expected from various dualities including
$SL(2,\bm{Z})$ duality in type IIB closed superstring theory
\cite{SchwarzSen},
and the corresponding classical solutions with Ramond-Ramond charges
were found in supergravity theory which describes the low energy
dynamics of superstring theory \cite{Schwarz}.
Then a microscopic construction of such ``solitonic'' objects was
given as hypersurfaces on which Dirichlet open strings end
\cite{DLP}.
Though first introduced as fixed hypersurfaces, D-branes become
dynamical objects with internal degrees of freedom originating from
Dirichlet open strings.
For example, a part of the massless gauge fields (``photons'') are
converted into the degrees of freedom representing the collective
coordinates for translation of D-branes.

The purpose of this paper is to introduce D-branes in covariant
bosonic string field theory (SFT), a formulation of string theory
as a straightforward extension of gauge field theories.
The most ambitious and interesting approach to D-branes using SFT
would be identify D-branes as
classical solutions (solitons) in closed SFT and carry out the
quantization around D-branes using the technique familiar in local
field theories \cite{Soliton}.
This attempt, however, seems very hard and might be impossible in view
of the $1/g$ dependence of the D-brane tension on the closed string
coupling constant $g$.
Instead, we shall adopt another way of introducing D-branes into
closed SFT. This is to add to the SFT action a term describing
the interaction between D-brane and closed string.
This interaction will be given as a product $B\cdot\Phi$ of closed
string field $\Phi$ and the ``boundary state'' $B$
\cite{ADGNSV,CLNY1,CLNY2,CLNY3,CallanKlebanov}.
The latter has been known to describe the initial (final) state of a
closed string emitted from (absorbed by) the D-brane.

Of course, we need a principle for introducing such a new
interaction. Our principle here is to keep the stringy local gauge
invariance present in the original closed SFT.
This stringy invariance includes, for example, the general coordinate
invariance and the gauge invariance associated with massless
anti-symmetric tensor field.
In order to preserve the gauge invariance after introducing the
$B\cdot\Phi$ interaction, the boundary state $B$ must also transform.
We shall see that this is realized by defining the transformation law
of the dynamical variables $\D$ associated with the boundary state.
The gauge invariance requirement also fixes the $\D$-dependence of the
normalization factor of the boundary state as well as the purely
$\D$-term which we have to add to the SFT action besides the
$B\cdot\Phi$ interaction.
In this paper we take as $\D$, the dynamical variable associated with
a D-brane, constant field strength $F_{\mu\nu}$ and the parameter
$\tlt_\mu^i$ representing the tilt of the D-brane.

Unfortunately, we do not yet have a fully satisfactory quantum
theory of covariant closed SFT. The origin of the problem is that,
although we have closed SFT actions having gauge invariance
\cite{HIKKOlettclosed,HIKKOclosed,Non-Poly},
their naive path-integral quantization leads to theories where
gauge invariance and unitarity are broken.
One way to remedy this defect is to add quantum corrections to
classical SFT action \cite{HataBV,Zwiebach} by following the
Batalin-Vilkovisky formalism \cite{BV}.
However, the resulting theories become too complicated to be used for
practical analysis.
In this paper, we ignore this quantization problem and construct a
``closed SFT + D-brane'' system on the basis of covariant closed SFT
proposed in refs.\ \cite{HIKKOlettclosed,HIKKOclosed}.
We hope that the essentials of this paper remain valid in a more
complete formulation.

Finally, we mention another way of describing D-branes in the
framework of SFT which we do not adopt in this paper.
This is to consider a field theory of Dirichlet open string.
Given a SFT for Neumann open string (e.g., the ones given in
\cite{WittenSFT,HIKKOopen}), the Dirichlet open SFT is obtained by
T-duality transformation \cite{GPR,KZ}: in the BRST charge and the
string vertices we have only to replace the center-of-mass momentum in
the transverse directions with the difference between the coordinates
of the two D-branes on which the open string end.
However, in this approach the closed string degrees of freedom are
treated rather indirectly since they appear dynamically as
loop effects in covariant open SFT.

The organization of the rest of this paper is as follows.
In Sec.\ \ref{sec:dLm}, we construct ``closed SFT + D-brane'' system
using the gauge invariance principle mentioned above.
The gauge transformation considered is the one which shifts the
anti-symmetric tensor field in $\Phi$ by a constant.
In Sec.\ \ref{sec:dLp}, we examine the gauge invariance under linear
coordinate transformation.
Though in Secs.\ \ref{sec:dLm} and \ref{sec:dLp} we take only
$F_{\mu\nu}$ as dynamical variable associated with D-brane, in
Sec.\ \ref{sec:tilt} we introduce the variable $\tlt_\mu^i$ specifying
the tilt of the D-brane. In Sec.\ \ref{sec:sigma}, the correspondence
between the $\sigma$-model approach and the present SFT approach is
studied. The final section is devoted to a summary and discussions.
In Appendix \ref{app:sft}, we summarize various formulas in SFT used
in the text, and in Appendix \ref{app:star}, we present the details
of the calculation of the star products used in Secs.\ \ref{sec:dLm},
\ref{sec:dLp} and \ref{sec:tilt}.

\clearpage
%%%%%%%%%%%%%%%%%%%%%%%%%%%%%%%%%%%%%%%%%%%%%%%%%%%%%%%%%%%%%%%%%%%%%%%%%
\section{Introducing D-brane to SFT}
\label{sec:dLm}

\subsection{Source term and gauge invariance}

We start with the system of closed SFT field $\Phi$
described by the action \cite{HIKKOclosed},
\begin{equation}
\Sz[\Phi]
=\frac{1}{g^2}\left\{\frac{1}{2}\,\Phi\cdot\QB\Phi
+ \frac{1}{3}\Phi\cdot(\Phi\star\Phi)\right\},
\label{eq:Sz}
\end{equation}
which  has an invariance under the stringy local gauge transformation
$\delta_\Lambda$,
\begin{eqnarray}
\delta_\Lambda\Phi = Q_B \Lambda +2\Phi\star\Lambda .
\label{eq:dL_Phi}
\end{eqnarray}
In eqs.\ (\ref{eq:Sz}) and (\ref{eq:dL_Phi}) the meaning of the
products $\cdot$ and $\star$ are as given in ref.\ \cite{HIKKOclosed}.
Compared with the closed string field $\Phi$ in ref.\
\cite{HIKKOclosed}, we have rescaled it by the coupling constant $g$
for later convenience.

We would like to extend this closed string field system
to the one containing the D-brane degrees of freedom.
As explained in Sec.\ \ref{sec:intro},
our principle of the extension is to keep the gauge invariance
(\ref{eq:dL_Phi}) intact.
Since D-brane can be regarded as a source of closed strings,
let us add to (\ref{eq:Sz}) the following source term
\begin{equation}
\Ss = B[\D]\cdot\Phi =\int\! dz_0\braket{B[\D]}{\Phi} .
\label{eq:Ss}
\end{equation}
Here, $\ket{B[\D]}$ represents the state for the emission and
absorption of closed strings. It is a function of new
dynamical degrees of freedom associated with the D-brane, which we
denote collectively by $\D$.
The integration measure $dz_0$ is over the zero-modes of the string
coordinates
$Z(\sigma)\equiv\left(X^M(\sigma),c(\sigma),\ac(\sigma)\right)$ and
the string-length parameter.
In the following we adopt the $\pi_c^0$-omitted formulation
\cite{HIKKOclosed} and the
representation $z_0\equiv(x^M,\ac_0,\wt{\alpha})$, where
$\wt{\alpha}$ is the variable conjugate to the string-length parameter
$\alpha$.\footnote{
String field $\Phi$ in SFT of refs.\ \cite{HIKKOclosed,HIKKOopen}
contains as its argument the string-length parameter $\alpha$ in
addition to $\left(X^M(\sigma),c(\sigma),\ac(\sigma)\right)$.
The $\wt{\alpha}$-representation is obtained from the
$\alpha$-representation by the Fourier transformation $\int\!d\alpha
\exp\left(i\alpha\wt{\alpha}\right)$:
$\alpha$ is a momentum-like variable while $\wt{\alpha}$ is a
coordinate-like one.
Physical quantities do not depend on $\wt{\alpha}$, and we have an
infinite number of equivalent worlds specified by $\wt{\alpha}$.
}
Since the string field $\Phi$ and the measure $dz_0$ have ghost number
$\NFP[\Phi]=-1$ and $\NFP[dz_0]=1$, respectively, $B[\D]$ should carry
$\NFP[B]=0$.
The Grassmann integration over $\ac_0$ is defined by
$\int\!d\ac_0\,\ac_0=1$.

Assuming that the transformation law (\ref{eq:dL_Phi}) of the closed
string field $\Phi$ remains unchanged after introducing $\Ss$,
the requirement that the system described by the action $\Sz+\Ss$
be invariant under the gauge transformation implies that the equation,
\begin{equation}
0=\dL\left(\Sz +\Ss\right)=\dL \Ss =
\dL B\cdot\Phi + B\cdot\left(\QB\Lambda+2\Phi\star\Lambda
\right) ,
\label{eq:dL_Ss=0}
\end{equation}
holds for any $\Phi$ and any $\Lambda$. This leads to the following
two conditions:
\begin{eqnarray}
&&\QB B[\D] =0 \label{eq:QB_B=0} ,\\
&&\dL B[\D] = 2 B[\D]\star\Lambda .
\label{eq:dL_B}
\end{eqnarray}
Namely, $B[\D]$ must be a BRST invariant state annihilated by $\QB$,
and the gauge transformation law $\dL\D$ must be determined so as to
satisfy the second condition (\ref{eq:dL_B}) (of course, there is no
guarantee at this stage that there exists $\dL\D$ satisfying
eq.\ (\ref{eq:dL_B})).
Since (\ref{eq:QB_B=0}) should hold for arbitrary $\D$, $\D$ can be
regarded as arbitrariness in the solution of $\QB B=0$ just like
the collective coordinates of soliton solutions in conventional field
theories.

Although in the complete treatment $\D$ is expected to represent the
same number of degrees of freedom as Dirichlet open strings,
we have not yet succeeded in developing a systematic
method of treating within our framework all the degrees of freedom
associated with D-brane.
In this section, we shall consider the simplest case of taking as $\D$
only the {\em constant} field strength $F_{\mu\nu}$ of the massless
gauge field on the D-$p$-brane. In Sec.\ \ref{sec:tilt} we shall add
to $\D$ the degrees of freedom representing the tilt of the D-brane.
Let $\mu=0,1,\cdots,p$ and $i=p+1,\cdots,d-1$
denote the space-time indices parallel and perpendicular to
the $p$-brane, respectively.
Then, the state $\ket{B(F)}$ satisfying the BRST-invariance
condition (\ref{eq:QB_B=0}) has been known as the boundary state
\cite{ADGNSV,CLNY1,CLNY2,CLNY3,CallanKlebanov}:
\begin{equation}
\ket{B(F)(x^M,\ac_0,\wt{\alpha})}=
N(F)\ket{\B{N}(F)}\otimes\ket{\B{D}}\otimes\ket{\B{gh}} ,
\label{eq:B}
\end{equation}
with the factor states $\ket{\B{N,D,gh}}$ given by\footnote{
The Lorentz indices $M=(\mu,i)$ are raised/lowered by using the flat
metric $\eta^{MN}=\eta_{MN}=(\eta_{\mu\nu},\delta_{ij})
={\rm diag}(-1,1,\cdots,1)$.
}
\begin{eqnarray}
&&\ket{\B{N}(F)}=\exp\left\{
-\sum_{n\ge 1}\frac{1}{n}\a_{-n}^{(+)\mu}
\calO(F)_\mu{}^\nu\a_{-n\,\nu}^{(-)}
\right\}\ket{0}_{p+1} ,
\label{eq:B_N}\\
&&\ket{\B{D}}=\exp\left\{
\sum_{n\ge 1}\frac{1}{n}\a_{-n}^{(+)\,i}
\a_{-n\,i}^{(-)}\right\}\ket{0}_{d-p-1}
\delta^{d-p-1}\bigl(x^i\bigr) ,
\label{eq:B_D}\\
&&\ket{\B{gh}}=\exp\left\{
\sum_{n\ge 1}\left(c_{-n}^{(+)}\ac_{-n}^{(-)}
+ c_{-n}^{(-)}\ac_{-n}^{(+)}\right)\right\}\ket{0}_{\rm gh} .
\label{eq:B_gh}
\end{eqnarray}
In eq.\ (\ref{eq:B_N}), $\calO$ is a $(p+1)\times(p+1)$ matrix
satisfying the orthonormality condition
\begin{equation}
\calO_\mu{}^\rho\eta_{\rho\lambda}\calO_{\nu}{}^\lambda
=\eta_{\mu\nu} ,
\label{eq:OetaO=O}
\end{equation}
and is expressed in terms of an anti-symmetric constant matrix
$F_{\mu\nu}$ as
\begin{equation}
\calO(F)_\mu{}^\nu\equiv
\left[\left(1 + F\right)^{-1}
\left(1 - F\right)\right]_\mu^{\;\nu} .
\label{eq:calO}
\end{equation}
The state $\ket{B(F)}$ is characterized by the
following conditions for the string coordinates,
\begin{eqnarray}
&&X^i(\sigma)\ket{B(F)}=0 ,
\label{eq:X_B=0}\\
&&\left(P_\mu(\sigma)-F_{\mu\nu}\Drv{}{\sigma}
X^\nu(\sigma)\right)\ket{B(F)}=0 ,
\label{eq:(P-FX)B=0}\\
&&\pi_c(\sigma)\ket{B(F)}=\pi_{\ac}(\sigma)\ket{B(F)}=0 .
\label{eq:pi_c_B=0}
\end{eqnarray}
They are equivalently expressed in terms of the oscillation modes as
\begin{eqnarray}
&&\left(\a^{(+)\,i}_{n} - \a^{(-)\,i}_{-n}\right)\ket{B(F)}=0 ,
\label{eq:X^i_B=0_oscl}\\
&&\left(\a^{(+)}_{n\,\mu}+\calO_\mu{}^\nu\a^{(-)}_{-n\,\nu}
\right)\ket{B(F)}
=\left(\a^{(-)\,\mu}_{n}+\a^{(+)\,\nu}_{-n}\calO_\nu{}^\mu
\right)\ket{B(F)}=0 ,
\label{eq:(P-FX)B=0_oscl}\\
&&\left(c^{(+)}_n + c^{(-)}_{-n}\right)\ket{B(F)}
=\left(\ac^{(+)}_n - \ac^{(-)}_{-n}\right)\ket{B(F)}=0 ,
\label{eq:pi_c_B=0_oscl}\\
&&\left(x^i, \Pdrv{}{x^\mu}, \Pdrv{}{\ac_0}\right)\ket{B(F)}=0 ,
\label{eq:zero-modes_B=0}
\end{eqnarray}
with $n=\pm 1, \pm 2, \cdots$.
The BRST invariance of $\ket{B(F)}$,
\begin{equation}
\QB\ket{B(F)}=0 ,
\label{eq:QB_B(F)=0}
\end{equation}
may be understood from the form of $\QB$ (see eq.\ (\ref{eq:QB_sc})
for the complete expression),
\begin{equation}
\QB=2\sqrt{\pi}\int_0^{2\pi}\!d\sigma\left\{
i\pi_{\ac}\left[\cdots\right]
-c\left(P_M X'^M + c'\pi_c + \pi'_{\ac}\ac\right)\right\} ,
\label{eq:QB_rough}
\end{equation}
and the properties (\ref{eq:X_B=0}), (\ref{eq:(P-FX)B=0}) and
(\ref{eq:pi_c_B=0}), and in particular, the anti-symmetric nature of
$F_{\mu\nu}$.\footnote{
Although the normal ordering for $\QB$ is ignored in this argument
based on eq.\ (\ref{eq:QB_rough}), the part of $\QB$ where the normal
ordering is relevant is $L\left(\p/\p\ac_0\right)$ (see
(\ref{eq:QB-II})), which annihilates $\ket{B(F)}$ since it is
independent of $\ac_0$.
}

The front factor $N(F)$ in eq.\ (\ref{eq:B}) cannot be determined at
this stage from the requirement of the BRST-invariance
(\ref{eq:QB_B(F)=0}) alone.
In the next subsection we shall fix $N(F)$ using the gauge invariance
requirement. The resultant $N(F)$ will agree with the one obtained
previously from quite a different argument \cite{CLNY3}.

\subsection{Determination of $N(F)$
and the transformation law of $F_{\mu\nu}$}

In this subsection we shall examine eq.\ (\ref{eq:dL_B}), the
condition to determine $\dL\D$. Since we have restricted $\D$ to a
subspace of {\em constat} field strength $F$, we are not allowed to
consider arbitrary gauge transformation functional $\Lambda$:
$\dL\D$ is in general not confined to our subspace.
As one of the ``allowed'' $\Lambda$,
let us consider the following $\Lambda_-$:
\begin{equation}
\ket{\Lambda_{-}(x,\ac_0,\wt{\alpha})}
=i\,\ac_0\left(
\a_{-1}^{(+)\mu}\ac_{-1}^{(-)}-\a_{-1}^{(-)\mu}\ac_{-1}^{(+)}
\right)\ket{0}\zeta_\mu(x,\wt{\alpha}) ,
\label{eq:Lambda_-}
\end{equation}
with $\zeta_\mu$ linear in $x^\mu$,
\begin{equation}
\zeta_\mu(x,\wt{\alpha})=a_{\mu\nu}x^\nu .
\label{eq:zeta_mu}
\end{equation}
The gauge transformation (\ref{eq:dL_Phi}) for this $\Lambda_-$
induces the shift
$\delta_{\Lambda_-}B_{\mu\nu}
=\p_\mu\zeta_\nu-\p_\nu\zeta_\mu + \ldots$
on massless anti-symmetric tensor $B_{\mu\nu}$ contained in
$\Phi$ (see Sec.\ \ref{sec:sigma}).

Since $\Lambda_-$ (\ref{eq:Lambda_-}) does not depend on
$\wt{\alpha}$, implying that it has vanishing string-length
$\alpha=0$, the star product $\Psi\star\Lambda_-$ for any string
functional $\Psi$ is expressed as
\begin{equation}
\ket{\Psi\star\Lambda_{-}}
=\Half\, a_{\mu\nu}\calD_{-}^{\mu\nu}\ket{\Psi} ,
\label{eq:Psi*L-}
\end{equation}
in terms of the linear anti-hermitian operator $\calD_{-}^{\mu\nu}$
given by
\begin{eqnarray}
&&\calD_{-}^{\mu\nu}=-i\int_0^{2\pi}\!\!d\sigma\,
\Drv{X^{\mu}(\sigma)}{\sigma}X^{\nu}(\sigma)
-\eta^{\mu\nu}\int_0^{2\pi}\!\!d\sigma\,
i\pi_c(\sigma)|_{\rm oscl}\cdot{}i\pi_{\ac}(\sigma)|_{\rm oscl}
\nn\\
&&=-\frac{1}{2}\sum_{n\ge 1}\frac{1}{n}\left(
\a_{-n}^{(+)\mu}\a_{n}^{(+)\nu}
-\a_{-n}^{(-)\mu}\a_{n}^{(-)\nu}
+\a_{n}^{(+)\mu}\a_{n}^{(-)\nu}
-\a_{-n}^{(+)\mu}\a_{-n}^{(-)\nu}
-[\mu\leftrightarrow\nu]
\right)
\nn\\
&&\qquad+ \mbox{(ghost coordinates part)} ,
\label{eq:Dminus}
\end{eqnarray}
where $i\pi_c(\sigma)\vert_{\rm oscl}$ denotes the non-zero mode part
of $i\pi_c(\sigma)$.
Details of the derivation of (\ref{eq:Psi*L-}) is presented in
Appendix \ref{app:star}.
The first term of $\calD_{-}^{\mu\nu}$,
$-i\int_0^{2\pi}d\sigma\left(dX^\mu/d\sigma\right)
\zeta_\mu\left(X\right)$,
is a geometrically natural one.
We do not know the intuitive interpretation of the ghost coordinate
part of $\calD_{-}^{\mu\nu}$.

Applying eq.\ (\ref{eq:Psi*L-}) for $\Psi=B(F)$, we obtain
\begin{eqnarray}
&&\ket{B(F)\star\Lambda_{-}}
=\Biggl\{-\Half\zzero\,\tr\left[\left(a-\tp{a}\right)\f\right]
\nn\\
&&\hspace*{3.5cm}
+\sum_{n\ge 1}\frac{1}{n}\a_{-n}^{(+)\mu}
\left[\f\left(a-\tp{a}\right)\f\right]_{\mu\nu}
\a_{-n}^{(-)\nu}\Biggr\}\ket{B(F)} ,
\label{eq:B*L-}
\end{eqnarray}
where $\tp{a}$ is the transposition of matrix $a$, and
$\zzero$ is the value of the zeta-function at the origin;
$\zzero=\sum_{n=1}^\infty 1$.
In deriving eq.\ (\ref{eq:B*L-}) we have used the fact that
$\ket{B(F)}$ is annihilated by the ghost coordinates part of
$\calD_{-}^{\mu\nu}$ due to (\ref{eq:pi_c_B=0}).

We would like to determine the transformation law $\dLm F$ of the
(constant) field strength $F$ under the present gauge transformation.
$\dLm F$ should satisfy (\ref{eq:dL_B}), namely,
\begin{equation}
\dLm\ket{B(F)}=2 \ket{B(F)\star\Lambda_-} .
\label{eq:dLm_B}
\end{equation}
It is easily seen that, if such a $\dLm F$ exists, it should be given
by
\begin{equation}
\dLm F_{\mu\nu}=\p_\nu\zeta_\mu-\p_\mu\zeta_\nu
=a_{\mu\nu}-a_{\nu\mu} ,
\label{eq:dLm_F}
\end{equation}
since the $\a_{-n}^{(+)}\a_{-n}^{(-)}$ term on the RHS of
(\ref{eq:B*L-}) is nothing but the variation of the exponent of
$\ket{\B{N}(F)}$ under this $\dLm$ (\ref{eq:dLm_F}).

In order for eq.\ (\ref{eq:dLm_B}) be satisfied completely,
the first term on the RHS of eq.\ (\ref{eq:B*L-}) must be  equal to
$\dLm\ln N(F)$:
\begin{equation}
\dLm\ln N(F)= - \zzero\,\tr\left[\left(a-\tp{a}\right)\f\right] ,
\label{eq:dLm_lnN(F)}
\end{equation}
This fixes $N(F)$ to be
\begin{equation}
N(F)=\frac{T_p}{4}\left[\det\left(1+F\right)\right]^{-\zzero} ,
\label{eq:N(F)}
\end{equation}
where $T_p$ is a constant, and the factor $1/4$ is for the convenience
of the comparison with the $\sigma$-model approach in Sec.\
\ref{sec:sigma}.
Thus we have determined $\dLm F$ as well as the front factor $N(F)$
which satisfy eq.\ (\ref{eq:dLm_B}).
The form of the front factor (\ref{eq:N(F)}) agrees with the one
determined by different arguments \cite{CLNY3}.

\subsection{Born-Infeld action}

One might think that the invariance of the system $\Sz+\Ss$
under the gauge transformation $\dLm$ has been established since we
have eqs.\ (\ref{eq:QB_B(F)=0}) and (\ref{eq:dLm_B}).  However, this
is not the case due to the fact that $\QB B(F)=0$ does not ensure
$B(F)\cdot\overrightarrow{\QB}\Lambda_-=0$ for the present $\Lambda_-$
(the right-arrow over $\QB$ indicates that it should operate on the
right). In fact, we have
\begin{equation}
B(F)\cdot\overrightarrow{\QB}\Lambda_-
=\int\!dz_0\bra{B(F)}\overrightarrow{\QB}\ket{\Lambda_-}
=2 V_{p+1}V_{\wt{\alpha}}\,
\tr\left[\left(a-\tp{a}\right)\f\right]N(F)
\ne 0 ,
\label{eq:B_QB_L-}
\end{equation}
where $V_{p+1}=\int\!d^{p+1}x$ is the space-time volume of the
D-$p$-brane and $V_{\wt{\alpha}}=\int\!d\wt{\alpha}$ is the
volume of the $\wt{\alpha}$-space.
The impossibility of reversing the direction of the arrow over $\QB$
in (\ref{eq:B_QB_L-}) may be explained as follows.
Note that the part of $\QB$ contributing to $\QB\ket{\Lambda_-}$ is
$i\sum_\pm
\left(c_{-1}^{(\pm)}\a_{1}^{(\pm)\nu}+c_{1}^{(\pm)}\a_{-1}^{(\pm)\nu}
\right)\!\p/\p x^\mu$.
Since $\Lambda_-\propto x$, we have, extracting the $x$-part of the
inner product,
\begin{equation}
B(F)\cdot\overrightarrow{\QB}\Lambda_-
\sim\int\! dx 1\frac{\overrightarrow\p}{\p x} x
\ne
-\int\! dx 1\frac{\overleftarrow\p}{\p x} x
\sim B(F)\cdot\overleftarrow{\QB}\Lambda_- =0 .
\label{eq:symbolic}
\end{equation}

To achieve the full gauge invariance under $\dLm$, we have to add
to $\Sz + \Ss$ a purely $F$ term $I(F)$ whose $\dLm$ transformation
cancels (\ref{eq:B_QB_L-}):
\begin{equation}
\dLm I(F) + B(F)\cdot\overrightarrow{\QB}\Lambda_- =0 .
\label{eq:dLm_I+B_QB_L-=0}
\end{equation}
In view of eq.\ (\ref{eq:dLm_lnN(F)}), the desired $I(F)$ satisfying
eq.\ (\ref{eq:dLm_I+B_QB_L-=0}) is seen to be given by
\begin{equation}
I(F)=\frac{2 V_{p+1}V_{\wt{\alpha}}}{\zzero}N(F)
=\frac{2}{\zzero}\int\! d^{p+1}x\!\int\! d\wt{\alpha}\,N(F) .
\label{eq:I(F)}
\end{equation}
This is nothing but the Born-Infeld action if we adopt the
zeta-function regularization $\zzero=-1/2$.

\subsection{Total action}

Summarizing the results of the previous subsections, the final
expression of our closed SFT system coupled to a D-brane is described
by the action $S_{\rm tot}[\Phi,F]$,
\begin{equation}
\S{tot}[\Phi,F]=\Sz[\Phi] + B(F)\cdot\Phi + I(F) .
\label{eq:Stot}
\end{equation}
$\S{tot}$ is invariant under the gauge transformation $\dLm$
with $\Lambda_-$ given by (\ref{eq:Lambda_-}) and (\ref{eq:zeta_mu}).
The transformation rules of $\Phi$ and $F$ are given respectively by
eqs.\ (\ref{eq:dL_Phi}) and (\ref{eq:dLm_F}).

%%%%%%%%%%%%%%%%%%%%%%%%%%%%%%%%%%%%%%%%%%%%%%%%%%%%%%%%%%%%%%%%%%%%%%%%%
\section{Linear coordinate transformation}
\label{sec:dLp}

As another gauge transformation which is closed within the constant
$F_{\mu\nu}$, let us consider the one corresponding to the linear
coordinate transformation,
$x^\mu\to x^\mu - \xi^\mu(x,\wt{\alpha})$ with
\begin{equation}
\xi^\mu(x,\wt{\alpha})=b^\mu{}_\nu\,x^\nu .
\label{eq:xi^mu}
\end{equation}
Such a coordinate transformation is generated by
\begin{equation}
\ket{\Lambda_{+}(x,\ac_0,\wt{\alpha})}
=i\,\ac_0\left(
\a_{-1\,\mu}^{(+)}\ac_{-1}^{(-)}+\a_{-1\,\mu}^{(-)}\ac_{-1}^{(+)}
\right)\ket{0}\xi^\mu(x,\wt{\alpha}) ,
\label{eq:Lambda_+}
\end{equation}
which is symmetric with respect to the left- and the right-moving
oscillators. Similarly to (\ref{eq:Psi*L-}), we have for any $\Psi$
(see Appendix \ref{app:star}),
\begin{equation}
\ket{\Psi\star\Lambda_{+}}
=\Half\, b_{\mu\nu}\calD_{+}^{\mu\nu}\ket{\Psi} ,
\label{eq:Psi*L+}
\end{equation}
where $\calD_{+}^{\mu\nu}$ is a linear operator given by
\begin{eqnarray}
&&\calD_{+}^{\mu\nu}=
\frac{i}{2}\int_0^{2\pi}\!\!d\sigma
\Bigl\{X^\nu(\sigma),P^\mu(\sigma)\Bigr\}
- \eta^{\mu\nu}\left(
\frac{1}{2}\left\{\wt{\alpha},\Pdrv{}{\wt{\alpha}}\right\}
+\G\right) .
\label{eq:Dplus}
\end{eqnarray}
On the RHS of eq.\ (\ref{eq:Dplus}), the oscillator expression of the
first term is
\begin{eqnarray}
&&\frac{i}{2}
\int_0^{2\pi}\!d\sigma\Bigl\{X^\nu(\sigma),P^\mu(\sigma)\Bigr\}
=\Half\left\{x^\nu,\Pdrv{}{x_\mu}\right\}
\nn\\
&&\qquad
-\Half\sum_{\pm}\sum_{n\ge 1}\frac{1}{n}\left(
\a_{-n}^{(\pm)\mu}\a_{n}^{(\pm)\nu}-\a_{-n}^{(\pm)\nu}\a_{n}^{(\pm)\mu}
+\a_{n}^{(\pm)\mu}\a_{n}^{(\mp)\nu}-\a_{-n}^{(\pm)\mu}\a_{-n}^{(\mp)\nu}
\right) ,
\label{eq:int_PX}
\end{eqnarray}
while the second term $\G$ consists solely of the ghost coordinates:
\begin{eqnarray}
&&\G=
\Half\int_0^{2\pi}\!\!d\sigma
\Bigl[\ac|_{\rm oscl}(\sigma), i\pi_{\ac}|_{\rm oscl}(\sigma)\Bigr]
+ \wtNFP
\nn\\
&&\phantom{\G}
=\Half\sum_\pm\sum_{n\ge 1}\left(
\ac_n^{(\pm)}c_n^{(\mp)}+\ac_{-n}^{(\pm)}c_{-n}^{(\mp)}
\right)+\Half\wtNFP ,
\label{eq:G}
\end{eqnarray}
with $\wtNFP$ being the oscillator part of the ghost
number operator (\ref{eq:wtNFP}).
Note that the first term of $\calD_{+}^{\mu\nu}$,
$(1/2)\int_0^{2\pi}d\sigma\left\{X^\nu,\delta/\delta X_\mu\right\}$,
is in fact the operator of linear coordinate transformation.
For $b^\mu{}_\nu$ with non-vanishing trace,
$b_{\mu\nu}\calD_{+}^{\mu\nu}$
also contains the $\wt{\alpha}$ and the ghost coordinate parts,
whose geometrical interpretation is not clear to us.

Applying eq.\ (\ref{eq:Psi*L+}) to $\Psi=B(F)$, we have
\begin{eqnarray}
&&2\ket{B(F)\star\Lambda_{+}}
=\Biggl\{\frac{1}{2}\,
\zzero\,\tr\left[\left(b+\tp{b}\right)\left(\F+1\right)\right]
\nn\\
&&\qquad\quad
+ 2\sum_{n\ge 1}\frac{1}{n}\a_{-n}^{(+)\mu}\left[
\frac{1}{1+F}\left(\tp{b}F + Fb\right)\frac{1}{1+F}\right]_{\mu\nu}
\a_{-n}^{(-)\nu}
\Biggr\}\ket{B(F)} .
\label{eq:B*L+}
\end{eqnarray}
In deriving eq.\ (\ref{eq:B*L+}), we have used in particular that
\begin{equation}
\G\ket{B(F)}= -\zzero\ket{B(F)} ,
\label{eq:G_B(F)}
\end{equation}
and that the zero-mode part of $\calD_{+}^{\mu\nu}$,
\begin{eqnarray}
&&\calD_+^{\mu\nu}\Big|_{\rm 0-modes}
=\Half\left\{x^\nu,\Pdrv{}{x_\mu}\right\}
-\Half\eta^{\mu\nu}\left\{\wt{\alpha},\Pdrv{}{\wt{\alpha}}\right\}
=x^\nu\Pdrv{}{x_\mu}-\eta^{\mu\nu}\wt{\alpha}\Pdrv{}{\wt{\alpha}} ,
\label{eq:calDplus_0}
\end{eqnarray}
annihilates $\ket{B(F)}$ since it depends on neither $x^\mu$ nor
$\wt{\alpha}$ for a constant $F$.

Our next task is to identify the transformation rule of $F$ under the
present gauge transformation $\dLp$. The equation for determining
$\dLp F$ is
\begin{equation}
\dLp\ket{B(F)}= 2 \ket{B(F)\star\Lambda_+} .
\label{eq:dLp_B}
\end{equation}
Since $\dLp$ corresponds to the linear coordinate transformation, it
is natural to take
\begin{equation}
\dLp F_{\mu\nu}=\p_\mu\xi^\lambda F_{\lambda\nu}
+ \p_\nu\xi^\lambda F_{\mu\lambda} + \xi^\lambda\p_\lambda F_{\mu\nu}
= b^\lambda{}_\mu F_{\lambda\nu}
+ b^\lambda{}_\nu F_{\mu\lambda} ,
\label{eq:dLp_F}
\end{equation}
or equivalently $\dLp F = \tp{b}\,F+F b$ in matrix notation.
However, since under (\ref{eq:dLp_F}) we have
\begin{eqnarray}
&&\dLp\ket{B(F)}=\Biggl\{
\Half\,\zzero\tr\left[
\left(b + \tp{b}\right)\left(\F -1\right)\right]
\nn\\
&&\hspace*{3cm}
+ 2\sum_{n\ge 1}\frac{1}{n}\a_{-n}^{(+)\mu}\left[
\frac{1}{1+F}\left(\tp{b}F + Fb\right)\frac{1}{1+F}\right]_{\mu\nu}
\a_{-n}^{(-)\nu}
\Biggr\}\ket{B(F)} ,
\label{eq:dLp_B(F)}
\end{eqnarray}
eq.\ (\ref{eq:dLp_B}) holds only for a traceless $b$ satisfying
$b^\mu{}_\mu=0$ due to disagreement between the oscillator independent
terms of eqs.\ (\ref{eq:B*L+}) and (\ref{eq:dLp_B(F)}).\footnote{
The tracelessness restriction $b^\mu{}_\mu=0$ persists even if we take
into account the change of the measure $dz_0$ since the variations of
$d^{p+1}x$ and $d\wt{\alpha}$ cancel each other as seen from
eq.\ (\ref{eq:calDplus_0}).
}
Note that the first term on the RHS of (\ref{eq:dLp_B(F)}) is the
contribution of $\dLp\ln N(F)$:
\begin{equation}
\dLp\ln N(F)=-\zzero\,\tr\left(\frac{1}{1+F}\,\dLp F\right)
=\Half\,\zzero\,\tr\left[
\left(b + \tp{b}\right)\left(
\F -1\right)\right] .
\label{eq:dLp_lnN(F)}
\end{equation}

To establish the invariance under $\dLp$, we have to confirm also
\begin{equation}
\dLp I(F) + B(F)\cdot\overrightarrow{\QB}\Lambda_+ =0 ,
\label{eq:dLp_I+B_QB_L+=0}
\end{equation}
for $I(F)$ of eq.\ (\ref{eq:I(F)}).
(Note that, since $\Lambda_+$ is proportional to $x$, reversing the
direction of the operation of $\QB$ in the inner product
$B(F)\cdot\overrightarrow{\QB}\Lambda_+$ is not allowed as in the case
of $\Lambda_-$.)
Using eq.\ (\ref{eq:dLp_lnN(F)}) and
\begin{equation}
B(F)\cdot\overrightarrow{\QB}\Lambda_+
=\int\!dz_0\bra{B(F)}\overrightarrow{\QB}\ket{\Lambda_+}
= - V_{p+1}\,V_{\wt{\alpha}}\,
\tr\left[\left(b+\tp{b}\right)\left(\F +1\right)\right]N(F) ,
\label{eq:B_QB_L+}
\end{equation}
we find that eq.\ (\ref{eq:dLp_I+B_QB_L+=0}) holds for traceless
$b$.

%%%%%%%%%%%%%%%%%%%%%%%%%%%%%%%%%%%%%%%%%%%%%%%%%%%%%%%%%%%%%%%%%%%%%%%%%
\section{Tilting the D-brane}
\label{sec:tilt}

So far we have considered a D-$p$-brane fixed at $x^i=0$
($i=p+1,\cdots,d-1$).
In this section we shall allow the D-brane to ``tilt'',
namely, consider a D-brane
\begin{equation}
x^i=\tlt_\mu^i x^\mu ,
\label{eq:tilted_D-brane}
\end{equation}
specified by $\tlt_\mu^i$ ($\tlt_\mu^i$ with $\mu\ne 0$ is the tilt
angle of the D-brane and $\tlt_0^i$ is its velocity).
The boundary state $\ket{B(F,\tlt)}$ corresponding to such a tilted
D-brane is obtained from $\ket{B(F)}$
of the previous sections as\footnote{
Note that $\ket{B(F)}$ itself can be expressed as
$\ket{B(F)}=\exp\left(
\frac{1}{2}F_{\mu\nu}\calD_{-}^{\mu\nu}\right)\ket{B(F=0)}$.
}
\begin{equation}
\ket{B(F,\tlt)}=U(\tlt)\ket{B(F)} ,
\label{eq:B=UB}
\end{equation}
where $U(\tlt)$ is a unitary operator,
\begin{equation}
U(\tlt)=\exp\left(-\tlt_\mu^i\calD_{+\,i}{}^\mu\right) ,
\label{eq:U}
\end{equation}
with $\calD_{+\,i}{}^\mu$ given by
\begin{equation}
\calD_{+\,i}{}^\mu = i\int_0^{2\pi}\!d\sigma
X^\mu(\sigma)P_i(\sigma) .
\label{eq:calD_t}
\end{equation}
In fact, $U(\tlt)$ effects the following transformation on the
string coordinates and their conjugates,
\begin{equation}
U(\tlt)\pmatrix{X^\mu\cr P_\mu\cr X^i\cr P_i} U(\tlt)^{-1}
=\pmatrix{X^\mu\cr P_\mu + \tlt_\mu^j P_j\cr
X^i - \tlt_\nu^i X^\nu\cr P_i} ,
\label{eq:U(P,X)U}
\end{equation}
and hence $\ket{B(F,\tlt)}$ (\ref{eq:B=UB}) satisfies,
instead of eqs.\ (\ref{eq:X_B=0}) and (\ref{eq:(P-FX)B=0}), the
following two:
\begin{eqnarray}
&&\left(X^i(\sigma)-\tlt_\nu^i X^\nu(\sigma)
\right)\ket{B(F,\tlt)}=0 ,
\label{eq:(X-fX)B=0}\\
&&\left(P_\mu(\sigma) + \tlt_\mu^j P_j(\sigma)
 - F_{\mu\nu}(dX^\nu(\sigma)/d\sigma)
\right)\ket{B(F,\tlt)}=0 .
\label{eq:(P+fP-FX')B=0}
\end{eqnarray}
Our new boundary state $\ket{B(F,\tlt)}$ is also BRST invariant:
\begin{equation}
\QB\ket{B(F,\tlt)}=0 .
\label{eq:QB_B(F,tlt)=0}
\end{equation}
This is easily seen by noting that $P_M(\sigma)X'^M(\sigma)$
contained in $\QB$ is invariant under the transformation of $U(\tlt)$
(recall eq.\ (\ref{eq:QB_rough})):
\begin{equation}
U(\tlt)\, P_M X'^M\,U(\tlt)^{-1}= P_M X'^M  .
\label{eq:UdXPU}
\end{equation}

We would like to repeat the construction of ``closed SFT + D-brane''
system of Sec.\ \ref{sec:dLm} by taking $F_{\mu\nu}$ {\em and}
$\tlt_\mu^i$ as dynamical variables associated with the D-brane.
The total action of this should be given by
\begin{equation}
\S{tot}[\Phi,F,\tlt]=\Sz[\Phi] + B(F,\tlt)\cdot\Phi + I(F,\tlt) ,
\label{eq:Stot(F,tlt)}
\end{equation}
and we shall determine $I(F,\tlt)$ and a possible
$(F,\tlt)$-dependent factor multiplied on $\ket{B(F,\tlt)}$
so that the system has gauge invariance as before.

Let us consider the gauge transformation $\dLm$.
$B(F,\tlt)$ and $I(F,\tlt)$ have to satisfy the following two
conditions:
\begin{eqnarray}
&&\dLm\ket{B(F,\tlt)}=2\ket{B(F,\tlt)\star\Lambda_-}
=a_{\mu\nu}\calD_{-}^{\mu\nu}\ket{B(F,\tlt)} ,
\label{eq:dLm_B(F,tlt)}\\
&&\dLm I(F,\tlt)+B(F,\tlt)\cdot\overrightarrow{\QB}\Lambda_- =0 .
\label{eq:dLm_I(F,tlt)+B(F,tlt)_QB_L-=0}
\end{eqnarray}
{}From (\ref{eq:dLm_B(F,tlt)}) we find that the transformation rule of
$F$ is as before and $\tlt$ is inert under $\dLm$,
\begin{equation}
\dLm F_{\mu\nu}= a_{\mu\nu} - a_{\nu\mu} ,
\quad
\dLm\tlt_\mu^i =0 ,
\label{eq:dLm_(F,tlt)}
\end{equation}
since $\calD_{-}^{\mu\nu}$ (\ref{eq:Dminus}) commutes with $U(\tlt)$,
$\left[U(\tlt), \calD_{-}^{\mu\nu}\right]=0$.
We need no extra $(F,\tlt)$-dependent factor multiplying
$\ket{B(F,\tlt)}$ of eq.\ (\ref{eq:B=UB}).

Our next task is the determination of $I(F,\tlt)$ satisfying
(\ref{eq:dLm_I(F,tlt)+B(F,tlt)_QB_L-=0}). We shall do this in a manner
different from Sec.\ \ref{sec:dLm}.
For this purpose, observe that
\begin{equation}
\ac_0\bra{0}a_{\mu\nu}\calD_{-}^{\mu\nu}\ket{B(F,\tlt)}
=-\Half\zzero\bra{\Lambda_-}\overleftarrow{\QB}\ket{B(F,\tlt)} .
\label{eq:Obs}
\end{equation}
This is easily understood by noticing that $\calD_-^{\mu\nu}$ and the
exponent of $\ket{B(F,\tlt)}$ are given as sums over the oscillator
level number $n$, and that each term in the $n$-summation in
$\calD_{-}^{\mu\nu}$  (\ref{eq:Dminus}) gives equal contribution to
the LHS of (\ref{eq:Obs}).
Then, comparing $\braket{0}{\,\mbox{eq.\ (\ref{eq:dLm_B(F,tlt)})}}$
and (\ref{eq:dLm_I(F,tlt)+B(F,tlt)_QB_L-=0}), we find that the desired
$I(F,\tlt)$ is given by
\begin{equation}
I(F,\tlt)=\frac{2}{\zzero}\int\! d^{d+1}x\int\! d\wt{\alpha}
\braket{0}{B(F,\tlt)} .
\label{eq:I=braket0B}
\end{equation}
To calculate the inner product
$\braket{0}{B(F,\tlt)}=\bra{0}U(\tlt)\ket{B(F)}$, let us express
$U(\tlt)$ (\ref{eq:U}) as
\begin{equation}
U(\tlt)=
e^{-i\tlt_\mu^i x^i p_\mu} e^{-u^\dagger(\tlt)} e^{u(\tlt)} ,
\label{eq:U=e^ue^u}
\end{equation}
with $u(\tlt)$ given by
(note that $\left[u(\tlt),u^\dagger(\tlt)\right]=0$)
\begin{equation}
u(\tlt)=\Half\,\tlt_\mu^i\sum_{n\ge 1}\frac{1}{n}
\left(\a_{-n}^{(+)} + \a_n^{(-)}\right)_i
\left(\a_n^{(+)} - \a_{-n}^{(-)}\right)^\mu .
\label{eq:u}
\end{equation}
Then, using eqs.\ (\ref{eq:X^i_B=0_oscl}) and
(\ref{eq:(P-FX)B=0_oscl}) to express the annihilation operators in
$u(\tlt)$ in terms of the creation ones and making use of the formula,
\begin{equation}
\bra{0}\exp\left(\Half a_a M_{ab} a_b\right)
\exp\left(\Half a_a^\dagger N_{ab} a_b^\dagger\right)\ket{0}
=\left[\det\left(1 - NM\right)\right]^{-1/2} ,
\label{eq:0expexp0}
\end{equation}
valid for creation/annihilation operators $(a_a^\dagger, a_a)$
with $[a_a,a_b^\dagger]=\delta_{a,b}$,
we obtain
\begin{equation}
\braket{0}{B(F,\tlt)}=
N(F)\left[\det\left(
\delta^\mu_{\,\nu} + \Half\left(
\eta^{\mu\lambda}+\calO^{\mu\lambda}\right)
\tlt_\lambda^i \tlt_\nu^i
\right)\right]^{-\zzero}\!\!
\delta^{d-p-1}\!\left(x^i-\tlt_\mu^i x^\mu\right) .
\end{equation}
Therefore, $I(F,\tlt)$ is given by
\begin{equation}
I(F,\tlt)=\frac{T_p}{2\zzero}\int\! d^{p+1}x\int\! d\wt{\alpha}
\left[-\det\left(\eta_{\mu\nu} + \tlt_\mu^i\tlt_\nu^i
+ F_{\mu\nu}\right)\right]^{-\zzero} .
\label{eq:I(F,tlt)}
\end{equation}
For the use in the next section, we also present a more explicit
expression of $\ket{B(F,\tlt)}$:
\begin{equation}
\ket{B(F,\tlt)}=N\left(F+\tlt\tlt\right)
\exp\left\{\sum_{n\ge 1}\frac{1}{n}\a_{-n}^{(+)M}A_{MN}\a_{-n}^{(-)N}
\right\}\ket{0}_{d}
\otimes\ket{\B{gh}}\delta^{d-p-1}\left(x^i-\tlt_\mu^i x^\mu\right) ,
\label{eq:B(F,tlt)}
\end{equation}
where $F+\tlt\tlt$ is short for $F_{\mu\nu}+\tlt_\mu^i\tlt_\nu^i$, and
the $d\times d$ matrix $A_{MN}$ is given by
\begin{equation}
\begin{array}{ll}
\ds A_{\mu\nu}=\eta_{\mu\nu}
-\left(2/\left(\wt{\eta}+F\right)\right)_{\mu\nu} ,
&\ds A_{\mu i}=
-\left(2/\left(\wt{\eta}+F\right)\right)_{\mu}^{\;\;\lambda}
\,\tlt_\lambda^i ,
\\[10pt]
\ds A_{i\mu}= -\tlt_\lambda^i
\left(2/\left(\wt{\eta}+F\right)\right)^\lambda_{\;\;\mu} ,
&\ds A_{ij}=\delta_{ij}-\tlt_\mu^i
\left(2/\left(\wt{\eta}+F\right)\right)^{\mu\nu}\tlt_\nu^j ,
\end{array}
\label{eq:A_MN}
\end{equation}
with
\begin{equation}
\wt{\eta}(\tlt)_{\mu\nu}=\eta_{\mu\nu} + \tlt_\mu^i\tlt_\nu^i .
\label{eq:wteta}
\end{equation}
$\ket{B(F,\tlt)}$ of (\ref{eq:B(F,tlt)}) has correct
normalization and satisfies the conditions (\ref{eq:(X-fX)B=0}) and
(\ref{eq:(P+fP-FX')B=0}).

Here, we have considered only the gauge transformation $\dLm$.
It is straightforward to confirm the invariances under other gauge
transformations closed within constant $F_{\mu\nu}$ and $\tlt_\mu^i$.
For example, under the transformation $\dLp$ of Sec.\ \ref{sec:dLp},
$F_{\mu\nu}$ and $\tlt_\mu^i$ should transform as
$\dLp F_{\mu\nu}=b^\lambda{}_\mu F_{\lambda\nu}
+b^\lambda{}_\nu F_{\mu\lambda}$
and $\dLp\tlt_\mu^i=b^\nu{}_\mu\tlt_\nu^i$, respectively.
Furthermore, the introduction of $\tlt_\mu^i$ allows the gauge
transformation $\dLt$ generated by
\begin{equation}
\ket{\Lambda_{t}}=i\,\ac_0\left(
\ac_{-1}^{(+)}\a_{-1\,i}^{(-)} +\ac_{-1}^{(-)}\a_{-1\,i}^{(+)}
\right)\ket{0}b_\mu^i x^\mu .
\label{eq:Lambda_t}
\end{equation}
For this $\Lambda_{t}$ we have
$\ket{\Psi\star\Lambda_{t}}
=\Half\tlt_\mu^i\calD_{+\,i}{}^\mu\ket{\Psi}$,
and the action $\S{tot}[\Phi,F,\tlt]$ (\ref{eq:Stot(F,tlt)}) is
invariant under $\dLt$ if the transformation law of
$(F_{\mu\nu},\tlt_\mu^i)$ is defined by
$\dLt F_{\mu\nu}=0$ and $\dLt\tlt_\mu^i=-b_\mu^i$.

%%%%%%%%%%%%%%%%%%%%%%%%%%%%%%%%%%%%%%%%%%%%%%%%%%%%%%%%%%%%%%%%%%%%%%%%%
\section{Comparison with the $\sigma$-model approach}
\label{sec:sigma}

In this section we shall examine the correspondence between our
``SFT + D-$p$-brane'' system (\ref{eq:Stot(F,tlt)}) and the
$\sigma$-model approach \cite{CLNY1,CLNY2,CLNY3}. We show that the
actions in both the approaches coincide with each other to the first
non-trivial orders in the expansion in powers of massless fields of
closed string.

First, let us consider the $\sigma$-model approach.
It has been known that the low energy dynamics of the system of closed
string coupled to a D-$p$-brane is described by the following effective
action \cite{PolchRev},
\begin{equation}
\S{eff}=\S{bulk} + \S{D} ,
\label{eq:S=S+S}
\end{equation}
with the bulk part $\S{bulk}$ and the D-brane action $\S{D}$ given
respectively by
\begin{eqnarray}
&&\S{bulk}=\frac{1}{g^2}
\int\! d^dx\sqrt{-G}e^{-2D}\left\{R + 4\left(\nabla D\right)^2
- \frac{1}{12}H^2\right\} ,
\label{eq:S_bulk}\\[5pt]
&&S_{\rm D}= - T_p\int\!d^{p+1}\sigma\, e^{-D}
\sqrt{-\det\left(
\wt{G}_{\mu\nu}+\wt{B}_{\mu\nu}+F_{\mu\nu}\right)} ,
\label{eq:S_D}
\end{eqnarray}
where $G_{MN}$ and $D$ are the metric and the dilaton fields,
respectively, and $H_{MNP}$ is the field strength of the
anti-symmetric tensor $B_{MN}$, $H=dB$.
In eq.\ (\ref{eq:S_D}), $\wt{G}_{\mu\nu}$ and $\wt{B}_{\mu\nu}$ are
the induced fields on the D-$p$-brane parameterized by the coordinate
$\sigma^\mu$ ($\mu=0,\cdots,p$).
Namely, letting $Y^M(\sigma)$ denote the D-brane space-time
coordinate, the induced metric $\wt{G}_{\mu\nu}(\sigma)$ is
\begin{equation}
\wt{G}_{\mu\nu}(\sigma)=
\p_\mu Y^M(\sigma)\,\p_\nu Y^N(\sigma)
\,G_{MN}\left(Y(\sigma)\right) .
\label{eq:wtG}
\end{equation}
The expression of $\wt{B}_{\mu\nu}$ is quite similar.

For comparing (\ref{eq:S=S+S}) with our SFT approach, let us make the
Weyl rescaling,
\begin{equation}
G_{MN}\to e^{4D/(d-2)}G_{MN} ,
\label{eq:Weyl}
\end{equation}
under which the bulk action (\ref{eq:S_bulk}) is reduced to
\begin{equation}
\S{bulk}=\frac{1}{g^2}\int\! d^dx\sqrt{-G}\left\{
R - \frac{4}{d-2}\left(\nabla D\right)^2
- \frac{1}{12}e^{-8D/(d-2)}H^2\right\} .
\label{eq:S_bulk_new}
\end{equation}
As for the D-brane action (\ref{eq:S_D}), we expanded it in powers of
the massless fields associated with closed string; $D$, $B_{MN}$  and
the metric fluctuation $h_{MN}$ defined for the Weyl rescaled
$G_{MN}$ by
\begin{equation}
G_{MN}=\eta_{MN}+ h_{MN} .
\label{eq:G=eta+h}
\end{equation}
Adopting the static gauge with $\sigma^\mu=x^\mu$, $Y^M(x)$ for the
tilted D-brane (\ref{eq:tilted_D-brane}) is
\begin{equation}
Y^\mu(x)=x^\mu,\quad Y^i(x)=\tlt_\mu^i x^\mu ,
\label{eq:Y}
\end{equation}
and hence the induced metric $\wt{G}_{\mu\nu}$ is given by
\begin{equation}
\wt{G}_{\mu\nu}=\wt{\eta}_{\mu\nu} + \wt{h}_{\mu\nu} ,
\label{eq:wtG=wteta+wth}
\end{equation}
in terms of $\wt{\eta}_{\mu\nu}$ of (\ref{eq:wteta}) and
$\wt{h}_{\mu\nu}$ defined by
\begin{equation}
\wt{h}_{\mu\nu}=h_{\mu\nu}+\tlt_\mu^i h_{i\nu}+\tlt_\nu^j h_{\mu j}
+ \tlt_\mu^i\tlt_\nu^j h_{ij} .
\label{eq:wth}
\end{equation}
The induced $\wt{B}_{\mu\nu}$ is also given by (\ref{eq:wth}) with $h$
replaced by $B$.
Then, keeping only the terms independent of and linear in
$\left(h_{MN}, D, B_{MN}\right)$, we have
\begin{equation}
\S{D}|_{\rm linear}
= - T_p\int\!d^{p+1}x\sqrt{-\det\left(\wt{\eta}+F\right)}
\left\{1 - D
+ \frac{2D}{d-2}\tr\!\left(\frac{\wt{\eta}}{\wt{\eta}+F}\right)
+ \Half\tr\!\left(\frac{\wt{h}+\wt{B}}{\wt{\eta}+F}\right)
\right\} .
\label{eq:S_D_exp}
\end{equation}

Next, let us consider our SFT approach.
The string field $\Phi$ consists of two parts, $\phi$ and $\psi$:
\begin{equation}
\ket{\Phi}= -\ac_0\ket{\phi} + \ket{\psi} .
\label{eq:Phi=phi+psi}
\end{equation}
They are expanded in terms of the component fields as
follows (we keep only the massless fields):
\begin{eqnarray}
&&\ket{\phi(x)}=\biggl\{%%\varphi (x) +
-\Half\hat{h}_{MN}(x)\left(\a^{M}_{-1}\a^{N}_{-1}\right)^{(+-)}
+\Half B_{MN}(x)\left(\a^{M}_{-1}\a^{N}_{-1}\right)^{[+-]} ,
\nn\\
&&\qquad\qquad
-\hat{D}(x)\left(c_{-1}\ac_{-1}\right)^{(+-)}
+f(x)\left(c_{-1}\ac_{-1}\right)^{[+-]}+\ldots \biggr\}\ket{0}
\label{eq:phi}\\[5pt]
&&\ket{\psi(x)}=\frac{i}{2}\left\{
b_M(x)\left(\a^M_{-1}\ac_{-1}\right)^{(+-)}
+e_M(x)\left(\a^M_{-1}\ac_{-1}\right)^{[+-]}
+\ldots \right\}\ket{0} ,
\label{eq:psi}
\end{eqnarray}
where we have used the (anti-)symmetrization symbol,
\begin{equation}
\left(ab\right)^{(+-)}\equiv
a^{(+)}b^{(-)}+a^{(-)}b^{(+)},\quad
\left(ab\right)^{[+-]}\equiv
a^{(+)}b^{(-)}-a^{(-)}b^{(+)} .
\label{eq:(anti)symm}
\end{equation}
In this section we omit the $\wt{\alpha}$-dependence of the fields.
Similarly, the component expansion of the gauge transformation
functional $\Lambda$ in (\ref{eq:dL_Phi}) is given as
\begin{eqnarray}
&&\ket{\Lambda}=i\ac_0\left\{
\xi_M(x)\left(\a_{-1}^M\ac_{-1}\right)^{(+-)}
+\zeta_M(x)\left(\a_{-1}^M\ac_{-1}\right)^{[+-]}+\ldots\right\}\ket{0}
\nn\\
&&\qquad\qquad
+\,\eta(x)\,\ac_{-1}^{(+)}\ac_{-1}^{(-)}\ket{0}
+ \ldots\ .
\label{eq:Lambda}
\end{eqnarray}

In the following we shall consider only the lowest non-trivial parts
of $\S{tot}$ (\ref{eq:Stot}) in the power series expansion in the
closed string massless component fields.
Therefore, in $S_0$ (\ref{eq:Sz}) we keep only the
kinetic term $(1/2g^2)\Phi\cdot\QB\Phi$.
Then, after integrating out the auxiliary fields $b_M$ and $e_M$ and
gauging $f$ away by using the gauge freedom of $\eta$ in
(\ref{eq:Lambda}), we obtain
\begin{equation}
\frac{1}{2g^2}\left.\Phi\cdot\QB\Phi\right|_{\rm massless}
=\frac{1}{g^2}\int\!d^dx\left\{
\left(\sqrt{-G}R\right)_{\rm quadratic}
- \frac{4}{d-2}\left(\p D\right)^2 - \frac{1}{12}H_{MNP}H^{MNP}
\right\} ,
\label{eq:PhiQBPhi}
\end{equation}
where we have reexpressed $\hat{h}_{MN}$ and $\hat{D}$ in
(\ref{eq:phi}) in terms of new $h_{MN}$ and $D$ as
\begin{equation}
\hat{h}_{MN}=h_{MN} +\frac{4}{d-2}D\,\eta_{MN},\quad
\hat{D}=\frac{4}{d-2}D + \Half h_M{}^M .
\label{eq:hat-nonhat}
\end{equation}
For the first term on the RHS of (\ref{eq:PhiQBPhi}), we have
used the formula
\begin{equation}
\left(\sqrt{-G}R\right)_{\rm quadratic}
=\frac{1}{4}h^{MN}\left(\BOX h_{MN}-2\p_N\p^P h_{MP}
+2\p_M\p_Nh^P_{\;P} -\eta_{MN}\BOX h^P_{\;P}\right) .
\label{eq:sqrtG_R}
\end{equation}
We see that eq.\ (\ref{eq:PhiQBPhi}) coincides with the part of
$\S{bulk}$ (\ref{eq:S_bulk_new}) quadratic in the fluctuations
$(h_{MN},D,B_{MN})$.

Then, let us consider the D-$p$-brane parts of $\S{tot}$
(\ref{eq:Stot(F,tlt)}).
Using eq.\ (\ref{eq:B(F,tlt)}) and keeping only the
massless component fields in $\Phi$, we obtain
\begin{eqnarray}
&&B(F,\tlt)\cdot\Phi|_{\rm massless}
=2\int\!d^{p+1}x N(F+\tlt\tlt)\left\{\Half\hat{h}_M{}^M
-\tr\!\left(\frac{\wt{\hat{h}}+\wt{B}}{\wt{\eta}+F}\right)
-\hat{D}\right\}
\nn\\
&&=-4\int\!d^{p+1}x\,N(F+\tlt\tlt)\left\{
- D + \frac{2D}{d-2}\tr\!\left(\frac{\wt{\eta}}{\wt{\eta}+F}\right)
+ \Half\tr\!\left(\frac{\wt{h}+\wt{B}}{\wt{\eta}+F}\right)
\right\} ,
\end{eqnarray}
where $\wt{\hat{h}}_{\mu\nu}$ is defined by (\ref{eq:wth}) with $h$
replaced by $\hat{h}$.
Adopting the zeta function regularization $\zzero=-1/2$, we find that
$B(F,\tlt)\cdot\Phi|_{\rm masssless} + I(F,\tlt)$ in SFT
approach indeed coincides with $\S{D}|_{\rm linear}$
(\ref{eq:S_D_exp}) for a common $T_p$.

Finally, we shall mention the gauge transformation properties of
the component fields.
The massless component fields appearing in (\ref{eq:PhiQBPhi})
transform under $\dL|_{\rm free}\Phi\equiv \QB\Lambda$ as
\begin{eqnarray}
&&\dL|_{\rm free} h_{MN}=\p_M\xi_N+\p_N\xi_M ,
\nn\\
&&\dL|_{\rm free} B_{MN}=\p_M\zeta_N-\p_N\zeta_M ,
\nn\\
&&\dL|_{\rm free} D=0 ,
\label{eq:dL_free}
\end{eqnarray}
and the free action (\ref{eq:PhiQBPhi}) is in fact invariant under
(\ref{eq:dL_free}). The transformation rule of the induced field
$\wt{B}_{\mu\nu}=B_{\mu\nu}+\tlt_\mu^i B_{i\nu}+\tlt_\nu^j B_{\mu j}
+\tlt_\mu^i\tlt_\nu^j B_{ij}$ under $\zeta_\mu$ of (\ref{eq:zeta_mu})
and $\zeta_i=0$ is
$\dLm\wt{B}_{\mu\nu}=\p_\mu\zeta_\nu-\p_\nu\zeta_\mu=
a_{\nu\mu}-a_{\mu\nu}$ as is expected from the fact that $\wt{B}$ and
$F_{\mu\nu}$ appear in $\S{D}$ (\ref{eq:S_D}) in the combination
$\wt{B}_{\mu\nu}+F_{\mu\nu}$.

We have seen the equivalence between our SFT approach and the
$\sigma$-model approach to the first non-trivial orders in the
expansion in powers of the massless closed string fields.
To discuss the equivalence to higher orders, we have to carry out the
integrations over the massive fields in our SFT approach.

%%%%%%%%%%%%%%%%%%%%%%%%%%%%%%%%%%%%%%%%%%%%%%%%%%%%%%%%%%%%%%%%%%%%%%%%%
\section{Summary and discussions}
\label{sec:summary}

We have constructed a system of closed SFT coupled to a D-brane
on the basis of gauge invariance principle.
Invariance under stringy local gauge transformation requires that the
state $B$ coupled to closed string field be annihilated by the BRST
charge $\QB$. The gauge invariance requirement also gives the equation
which determines the transformation law of the dynamical variable $\D$
associated with the D-brane.
Adopting as $B$ the boundary state $B(F,\tlt)$ which is a BRST invariant
one, invariance requirement under a special gauge
transformation which shifts the anti-symmetric tensor by a constant
fixes the $(F,\tlt)$-dependence of the front factor of $B$ as well as
the gauge transformation law of the field strength $F_{\mu\nu}$
and the tilt angle $\tlt_\mu^i$ of the D-brane. Furthermore, due to
the unboundedness of this gauge transformation at infinity, we need to
introduce the Born-Infeld action to realize the gauge invariance.
The invariance under linear coordinate transformation has also been
studied.
We have checked the correspondence between the action of our ``closed
SFT + D-brane'' system and the effective action in the $\sigma$-model
approach.

Our construction here is still incomplete and there are many subjects
to be studied.
One of the most important among them is to extend the dynamical
variables associate with D-brane. In this paper, we considered only
{\em constant} $F_{\mu\nu}$ and $\tlt_\mu^i$.
Since we know that D-branes have the same number of degrees of freedom
as Dirichlet open string, we should be able to incorporate all of them
in the present formalism. This extension includes generalizing
constant $(F_{\mu\nu},\tlt_\mu^i)$ to $x^\mu$-dependent ones
$(A_\mu(x),Y^i(x))$,
as well as introducing massive degrees of freedom on D-branes.
One way of introducing non-constant $F_{\mu\nu}$ is to
make use of gauge transformation.
Assuming that the gauge transformation (\ref{eq:Lambda_-})
with a general $\zeta_\mu(x)$ generates
$\dLm F_{\mu\nu}(x)=\p_\mu\zeta_\nu(x)-\p_\nu\zeta_\mu(x)$,
we can determine the boundary state for an (infinitesimally)
non-constant field strength $F_{\rm new}(x)=F + d\zeta(x)$ as
$B(F_{\rm new})=B(F) + 2B(F)\star\Lambda_-$.
Details of this extension will be given in a separate paper
\cite{HH2}.

As another problem left in our formalism, we have to determine
the D-brane tension $T_p$.
In the world sheet approach,
the D-brane tension has been determined by either using
Lovelace-Fischler-Susskind mechanism \cite{Lovelace,FS} or by
comparing the one-loop vacuum energy of Dirichlet open string with the
amplitude of massless field exchange in the low energy effective
action $\S{eff}$ (\ref{eq:S=S+S}) \cite{PolchRev}.
In our SFT approach, the boundary state $B$ has been introduced
as a state satisfying the BRST invariance condition (\ref{eq:QB_B=0}),
which is a linear equation and does not fix the absolute
magnitude of $B$.
It would be most interesting if we could ``improve'' our formalism in
such a way that the boundary state is determined by a non-linear
equation which allows the interpretation of the D-brane as a soliton
in closed SFT.

Although we do not know how to determine the absolute value of $T_p$
within our formalism, its dependence of the string coupling constant
$g$ can be deduced from the relation between the string coupling
constant and the dilaton expectation value. This well-known relation
is expressed in closed SFT as the property that $\Sz$ (\ref{eq:Sz}) is
invariant under following transformation of the string field $\Phi$ and
the coupling constant $g$
\cite{Yoneya,HataNagoshi,KZ,Kawano,HataDilaton}:
\begin{eqnarray}
&&\dD\ket{\Phi} =\left(\calD+\frac{d-2}{2}\right)\ket{\Phi}
+ 2\sqrt{d-2}\ket{\mbox{Dilaton}} ,
\label{eq:dD_Phi}\\
&&\dD g = \frac{d-2}{2}\,g ,
\label{eq:dD_g}
\end{eqnarray}
where $\calD$ is the dilatation operator defined by eq.\ (6)
of ref.\ \cite{HataDilaton}, and $\ket{\mbox{Dilaton}}$ is the zero
momentum dilaton state. In our ``closed SFT + D-brane'' system,
we can show that the total action $\S{tot}$ of eqs.\ (\ref{eq:Stot})
and (\ref{eq:Stot(F,tlt)}) is invariant under $\dD$ of eqs.\
(\ref{eq:dD_Phi}) and (\ref{eq:dD_g}) and suitably defined
$\dD\!\left(F_{\mu\nu},\tlt_\mu^i\right)$, if $\dD T_p$ is given by
\begin{equation}
\dD T_p= -\frac{d-2}{2}T_p .
\label{eq:dD_T_p}
\end{equation}
Eqs.\ (\ref{eq:dD_g}) and (\ref{eq:dD_T_p}) implies that
$T_p\propto 1/g$.

%%%%%%%%%%%%%%%%%%%%%%%%%%%%%%%%%%%%%%%%%%%%%%%%%%%%%%%%%%%%%
\vspace{.7cm}
\noindent
{\Large\bf Acknowledgments}\\[.2cm]
We would like to thank T.\ Kugo and T.\ Takahashi for valuable
discussions.

\newpage
%%%%%%%%%%%%%%%%%%%%%%%%%%%%%%%%%%%%%%%%%%%%%%%%%%%%%%%%%%%%%
\vspace{1.5cm}
\centerline{\Large\bf Appendix}
\appendix

\section{Summary of the formulas in closed SFT}
\label{app:sft}

In this appendix we summarize various quantities in closed SFT used in
the text.

\noindent
\underline{String coordinates}

String coordinates $\left(X^M(\sigma),c(\sigma),\ac(\sigma)\right)$
and their conjugates
$\left(P_M(\sigma),\pi_c(\sigma),\pi_{\ac}(\sigma)\right)$
are expanded in terms of the creation/annihilation operators as
follows.
For the space-time coordinate, we have
\begin{eqnarray}
&&X^M(\sigma)=\isqrt{\pi}\left\{ x^M +
\frac{i}{2}\sum_{n\ne 0}\frac{1}{n}\left(
\a_n^{(+)M} - \a_{-n}^{(-)M}\right)e^{in\sigma}\right\} ,
\label{eq:X}\\
&&P_M(\sigma)=-i\frac{\delta}{\delta X^\mu(\sigma)}
=\frac{1}{2\sqrt{\pi}}\left\{p_M + \sum_{n\ne 0}\left(
\a_n^{(+)M} + \a_{-n}^{(-)M}\right)e^{in\sigma}\right\} ,
\label{eq:P}
\end{eqnarray}
with $p_M=-i\pdrv{}{x^M}$.
For the ghost coordinates,
\begin{eqnarray}
&&\ac(\sigma)=\frac{1}{2\sqrt{\pi}}\left\{
\ac_0 + \sum_{n\ne 0}\left(
\ac_n^{(+)}+\ac_{-n}^{(-)}\right)e^{in\sigma}\right\} ,
\label{eq:ac}\\
&&i\pi_{\ac}(\sigma)=\frac{\delta}{\delta \ac(\sigma)}
=\frac{1}{2\sqrt{\pi}}\left\{
2\Pdrv{}{\ac_0} + \sum_{n\ne 0}\left(
c_n^{(+)}+c_{-n}^{(-)}\right)e^{in\sigma}\right\} ,
\label{eq:pi_ac}\\
&&c(\sigma)=-\frac{1}{2\sqrt{\pi}}\left\{
i\Pdrv{}{\pi_c^0}+\sum_{n\ne 0}\left(
c_n^{(+)} - c_{-n}^{(-)}\right)e^{in\sigma}\right\} ,
\label{eq:c}\\
&&i\pi_c(\sigma)=\frac{\delta}{\delta c(\sigma)}
=-\frac{1}{2\sqrt{\pi}}\left\{-2i\pi_c^0
+\sum_{n\ne 0}\left(\ac_n^{(+)}-\ac_{-n}^{(-)}\right)e^{in\sigma}
\right\} .
\label{eq:pi_c}
\end{eqnarray}
The (anti-)commutation relations among
$\left(\a_n^{(\pm)M},c_n^{(\pm)},\ac_n^{(\pm)}\right)$ ($n\ne 0$)
are as follows:
\begin{equation}
\left[\a_n^{(\pm)M},\a_m^{(\pm)N}\right]
=n\delta_{n+m,0}\eta^{MN},
\quad
\left\{c_n^{(\pm)},\ac_m^{(\pm)}\right\}=\delta_{n+m,0},
\quad\mbox{others}=0 .
\label{eq:CR}
\end{equation}

\noindent
\underline{BRST charge $\QB$}

The BRST charge in terms of the string coordinates and their
conjugates is given (modulo normal ordering) by
\begin{eqnarray}
&&\QB=2\sqrt{\pi}\int_0^{2\pi}\!d\sigma\Biggl\{i\pi_{\ac}\left[
-\Half\left(\eta^{MN}P_M P_N
+\eta_{MN}X'^{M}X'^{N}\right)
+i\left(c'\ac'-\pi'_{\ac}\pi_c\right)\right]
\nn\\
&&\hspace*{4cm}
-c\left(P_M X'^M + c'\pi_c + \pi'_{\ac}\ac\right)\Biggr\} ,
\label{eq:QB_sc}
\end{eqnarray}
where the prime denotes the differentiation with respect to $\sigma$.
In terms of the creation and the annihilation operators,
$\QB$ in the $\pi_{\ac}^0$-omitted formalism is
\begin{equation}
\QB=-\sum_\pm\sum_{n}: c^{(\pm)}_{-n}\left(
\sum_m\Bigl[\a^{(\pm)}_{n-m}\cdot\a^{(\pm)}_m
+ (n+m)\ac^{(\pm)}_{n-m}c^{(\pm)}_m\Bigr] - 2\delta_{n,0}
\right)\!:\ ,
\label{eq:QB}
\end{equation}
with $\alpha^{(\pm)}_{0\,\mu}=p_\mu/2$ and $c_0=\pdrv{}{\ac_0}$.
Making the dependence on the ghost zero-mode $\ac_0$ explicit, $\QB$
is also expressed as
\begin{equation}
\QB=L\Pdrv{}{\ac_0}+ \wtQB + \ac_0 M ,
\label{eq:QB-II}
\end{equation}
where $L$ and $M$ are given by
\begin{eqnarray}
&&L=-\Half p^2 -2\sum_\pm
\sum_{n\ge 1}\left[
\a_{-n}^{(\pm)}\cdot\a_n^{(\pm)}
+ n\left(c_{-n}^{(\pm)}\ac_{n}^{(\pm)}
+ \ac_{-n}^{(\pm)}c_{n}^{(\pm)}\right)\right] + 4 ,
\label{eq:L}\\
&&M=2\sum_\pm\sum_{n\ge 1}n c^{(\pm)}_{-n}c^{(\pm)}_{n} ,
\label{eq:M}
\end{eqnarray}
and $\wtQB$ is the part of $\QB$ (\ref{eq:QB}) containing neither
$\ac_0$ nor $c_0$.
$\wtQB$ contains, in particular, the part
\begin{equation}
\wtQB=i\sum_\pm\left(
c_{-1}^{(\pm)}\a_1^{(\pm)\mu}+c_{1}^{(\pm)}\a_{-1}^{(\pm)\mu}
\right)\Pdrv{}{x^\mu} + \ldots\ ,
\label{eq:wtQB=...}
\end{equation}
which contributes to $B(F)\cdot\QB\Lambda_\pm$.

\noindent
\underline{Ghost number operator}

The ghost number operator $\NFP{}$ is given by
\begin{equation}
\NFP=\ac_0\Pdrv{}{\ac_0} + \wtNFP ,
\label{eq:NFP}
\end{equation}
with
\begin{equation}
\wtNFP=\sum_\pm\sum_{n\ge 1}\left(
c_{-n}^{(\pm)}\ac_{n}^{(\pm)} -\ac_{-n}^{(\pm)}c_{n}^{(\pm)}
\right) .
\label{eq:wtNFP}
\end{equation}

\noindent
\underline{Dot and star products}

For general string functionals $\Phi_i$ ($i=1,2,3$), the dot and the
star products satisfy the following properties:
\begin{eqnarray}
&&\Phi_1\cdot\Phi_2 = (-)^{\abs{1}\abs{2}}\Phi_2\cdot\Phi_1 ,
\label{eq:prod1}\\
&&\Phi_1\star\Phi_2= -(-)^{\abs{1}\abs{2}}\,\Phi_2\star\Phi_1 ,
\label{eq:prod2}\\
&&\Phi_1\cdot\left(\Phi_2\star\Phi_3\right)
=(-)^{\abs{1}\left(\abs{2}+\abs{3}\right)}
\,\Phi_2\cdot\left(\Phi_3\star\Phi_1\right) ,
\label{eq:prod3}\\
&&\Phi_1\cdot\QB\Phi_2=-(-)^{\abs{1}}\left(\QB\Phi_1\right)\cdot\Phi_2 ,
\label{eq:prod4}
\end{eqnarray}
where $\abs{i}$ ($i=1,2,3$) is $0$ ($1$) if $\Phi_i$ is Grassmann-even
(-odd).

%%%%%%%%%%%%%%%%%%%%%%%%%%%%%%%%%%%%%%%%%%%%%%%%%%%%%%%%%%%%%%%%%%%
\section{Derivation of $\Psi\star\Lambda_{\pm}$}
\label{app:star}

In this appendix, we present the details of the calculation of
$\Psi\star\Lambda_{\pm}$, eqs.\ (\ref{eq:Psi*L-}) and
(\ref{eq:Psi*L+}).
The following calculations are based on the $(p_\mu,\alpha)$
representation, which is the Fourier transform of the
$(x^\mu,\wt{\alpha})$ representation adopted in the text.
Therefore, the gauge transformation functional $\Lambda_\pm$
we have to consider here are given in bra form by
\begin{equation}
\bra{\Lambda_\pm(p_\mu,\ac_0,\alpha)}=i\ac_0 \bra{0}\left(
\a_{1\,\mu}^{(+)}\ac_1^{(-)}\pm \a_{1\,\mu}^{(-)}\ac_1^{(+)}\right)
\e_\pm^\mu(p)\, \delta(\alpha)\times (2\pi)^{d+1} ,
\label{eq:Lambda_pm}
\end{equation}
where $\e_+^\mu(p)$ and $\e_-^\mu(p)$ are the Fourier transforms of
$\zeta^\mu(x)$ (\ref{eq:zeta_mu}) and $\xi^\mu(x)$ (\ref{eq:xi^mu}),
respectively:
\begin{equation}
\e_+^\mu(p)=ib^\mu{}_\nu\Pdrv{}{p_\nu}\delta^d(p) ,
\quad
\e_-^\mu(p)=ia^\mu{}_\nu\Pdrv{}{p_\nu}\delta^d(p) .
\label{eq:e_pm}
\end{equation}
The closed SFT three string vertex $\ket{V(1,2,3)}$ in the
$\pi_c^0$-omitted formulation is
\begin{eqnarray}
&&\ket{V(1,2,3)}=\left[\mu(1,2,3)\right]^2
\wp^{(1)}\wp^{(2)}\wp^{(3)} \nn\\
&&\phantom{\ket{V(1,2,3)}=}\times
\prod_{r=1}^3\left(1 - \ac_0^{(r)}\frac{1}{\sqrt{2}}w_I^{(r)}\right)
\exp\left(F(1,2,3)\right)\ket{0}_{1,2,3} \delta(1,2,3) ,
\label{eq:V(1,2,3)}
\end{eqnarray}
with $F(1,2,3)$ and $\delta(1,2,3)$ given by
\begin{eqnarray}
&&F(1,2,3)=\sum_\pm\sum_{r,s}\sum_{n,m\ge 1}\N{nm}{rs}\left(
\Half\a_{-n}^{(\pm)(r)}\cdot\a_{-m}^{(\pm)(s)}+
i\gamma_{-n}^{(\pm)(r)}\overline{\gamma}_{-m}^{(\pm)(s)}
\right)\nn\\
&&\phantom{F(1,2,3)=}
+\Half\sum_\pm\sum_r\sum_{n\ge 1}\N{n}{r}\a_{-n}^{(\pm)(r)}
\cdot\P + \tau_0\sum_{r=1}^3\frac{1}{\alpha_r}\frac{p_r^2}{4} ,
\label{eq:F(1,2,3)}\\[5pt]
&&\delta(1,2,3)=(2\pi)^{d}\delta^{d}
\Bigl(\sum_{r=1}^3 p_r\Bigr)
\cdot 2\pi\delta\Bigl(\sum_{r=1}^3
\alpha_r\Bigr) .
\label{eq:delta(1,2,3)}
\end{eqnarray}
Definitions of various quantities appearing in
eqs.\ (\ref{eq:V(1,2,3)}), (\ref{eq:F(1,2,3)}) and
(\ref{eq:delta(1,2,3)}) are found in \cite{HIKKOclosed,HIKKOpre}
(see also \cite{HataNagoshi}).
Some of them are given below when we use their explicit expressions.
For later convenience we divide $F(1,2,3)$ (\ref{eq:F(1,2,3)}) into
two parts:
\begin{eqnarray}
&&F=F_{\rm oscl} + F_{p^2} ,\qquad
F_{p^2}\equiv
\tau_0\sum_{r=1}^3\frac{1}{\alpha_r}\frac{p_r^2}{4} .
\label{eq:F=F+F}
\end{eqnarray}

Now, what we have to calculate is
\begin{eqnarray}
&&\int\!d1 \braket{\Lambda_\pm(1)}{V(1,2,3)}=\wp^{(2)}\wp^{(3)}
\lim_{\e\to 0}\lim_{p_1\to 0}\pmatrix{b_{\mu\nu}\cr a_{\mu\nu}}
\Pdrv{}{p_1^\nu}\ket{A_\pm^\mu(1,2,3)}_{2,3} ,
\label{eq:ToBeCalc}
\end{eqnarray}
with $\ket{A_\pm^\mu(1,2,3)}_{2,3}$ given by
\begin{eqnarray}
&&\ket{A_\pm^\mu(1,2,3)}_{2,3}\equiv
\left[\mu(1,2,3)\right]^2
\cdot (2\pi)^{d+1}\delta^{d}\Bigl(p_1+\sum_{s=2,3}p_s\Bigr)
\delta\Bigl(\e +\sum_{s=2,3}\alpha_s\Bigr)
\nn\\
&&\times{}_1\kern-3pt\bra{0}\left(
\a_1^{(+)\mu} \ac_1^{(-)} \pm \a_1^{(-)\mu} \ac_1^{(+)}\right)
\prod_{r=2,3}
\left(1 - \ac_0^{(r)}\frac{1}{\sqrt{2}}w_I^{(r)}\right)
\exp\left(F(1,2,3)\right)\ket{0}_{1,2,3}\ .
\label{eq:A}
\end{eqnarray}
Here, we have used the abbreviation
$d1\equiv d\ac_0^{(1)}d^{d}p_1/(2\pi)^{d}\cdot{}d\alpha_1/2\pi$,
and $\e$ is the string-length of the string 1, $\e\equiv\alpha_1$.

\noindent
\underline{Expansion of the quantities in $V$}

Eq.\ (\ref{eq:ToBeCalc}) tells that we have to take the $\e\to 0$
limit of the part of $\ket{A_\pm^\mu(1,2,3)}_{2,3}$
proportional to $p_1^\nu$.
For this purpose, we Laurent-expand various quantities in the
vertex $V(1,2,3)$ with respect to $\e/\alpha_2$.
First, we have (c.f. Appendix A of ref.\ \cite{HataNagoshi})
\begin{eqnarray}
&&\tau_0\equiv\sum_r \alpha_r\ln\abs{\alpha_r}
=\e\left\{
\ln\abs{\IRT{2}} - \frac{\e}{2\alpha_{2}}
+O(\e^2)\right\} ,
\label{eq:tau-exp}\\
&&\left[\mu(1,2,3)\right]^2 \equiv \exp\left(
-2\tau_0\sum_{r=1}^3\frac{1}{\alpha_r}\right)
=\abs{\RT{2}}^2\left(1 + \frac{\e}{\alpha_{2}} +O(\e^2)\right) ,
\label{eq:mu-exp}\\
&&\sum_{r=1}^3\frac{1}{\alpha_r}\frac{p_r^2}{4}
=\frac{1}{4\e}\left(
p_1^2 -p_1\!\cdot\!(p_1+2p_2)\,\frac{\e}{\alpha_2}
+O(\e^2) \right) ,
\label{eq:F_p^2/tau-exp}
\end{eqnarray}
where $O(\e)$ represents precisely $O(\e/\alpha_{2})$.
Therefore, $F_{p^2}$ of (\ref{eq:F=F+F}) is expanded as
\begin{eqnarray}
&&F_{p^2}=-\frac{1}{4}\left(p_1\right)^2\ln\abs{\RT{2}}
+O\!\left(\e p_1\ln\abs{\RT{2}}\right)
+ O\!\left(\e\left(p_1\right)^2\right)
+ O\!\left(\e^2\right) .
\label{eq:F_p^2-exp}
\end{eqnarray}
Next, let us expand the quantities associated with the
creation/annihilation operators.
For the Neumann coefficients, we have
\begin{eqnarray}
&&\N{1}{1}=\frac{1}{\e}e^{\tau_0/\e}
=\frac{\sgn(\e\alpha_{2})}{e\alpha_{2}}
\left(1 - \frac{\e}{2\alpha_{2}}
+ O(\e^2)\right) ,
\label{eq:N_1^1}\\
&&\N{m}{s}=\frac{1}{m\alpha_2}\times\cases{1& $(s=2)$\cr (-)^m&$(s=3)$}
+\ldots\ ,
\label{eq:N_m^s}\\
&&\N{1m}{1s}=
\abs{\IRT{2}}\times\cases{1& $(s=2)$\cr -(-)^m&$(s=3)$}
+ \ldots\ ,
\label{eq:N_1m^1s}\\
&&\N{nm}{23}= -\frac{(-)^n}{n}\delta_{n,m}\left(1+O(\e^2)\right)
-\left(1-\delta_{n,m}\right)\frac{(-)^n}{n-m}\frac{\e}{\alpha_2}
+\ldots\ ,
\label{eq:N_23}
\end{eqnarray}
where the dots $\ldots$ represents terms of higher order in
$\e/\alpha_2$.
Using (\ref{eq:N_m^s}) and (\ref{eq:N_23}), $F_{\rm oscl}$ with the
creation operators of the string 1 set equal to zero is expanded as
follows:
\begin{eqnarray}
&&F_{\rm oscl}(1,2,3)\Big\vert_{\mbox{\scriptsize oscl}^{(1)}\equiv 0}
=-\sum_\pm\sum_{n\ge 1}(-)^n\left(
\frac{1}{n}\a_{-n}^{(\pm)(2)}\cdot\a_{-n}^{(\pm)(3)}
- c_{-n}^{(\pm)(2)}\ac_{-n}^{(\pm)(3)}
+ \ac_{-n}^{(\pm)(2)}c_{-n}^{(\pm)(3)}\right)\nn\\
&&%%\hspace{4.5cm}
-\Half\sum_{\pm}\sum_{n\ge 1}\frac{1}{n}
\left(\a_{-n}^{(\pm)(2)}+(-)^n\a_{-n}^{(\pm)(3)}\right)\cdot p_1
-\frac{\e}{\alpha_2}\sum_\pm\sum_{n\ge 1}(-)^n\left(
c_{-n}^{(\pm)(2)}\ac_{-n}^{(\pm)(3)}
-c_{-n}^{(\pm)(3)}\ac_{-n}^{(\pm)(2)}
\right)
\nn\\
&& + \frac{\e}{\alpha_2}
\times\left(\mbox{$\alpha_{-n}^{(2)}\cdot\alpha_{-m}^{(3)}$ and
$c_{-n}^{(2,3)}\ac_{-m}^{(3,2)}$ with $n\ne m$}\right)
+ O(\e^2) .
\label{eq:F_a=0}
\end{eqnarray}
The fourth term on the RHS of (\ref{eq:F_a=0}) does not contribute to
our final result due to the projector $\wp$ and hence will be omitted
hereafter. Then, from (\ref{eq:F_a=0}) and (\ref{eq:F_p^2-exp}), we
obtain
\begin{eqnarray}
&&\left[\mu(1,2,3)\right]^2 \exp\left(
F_{\rm oscl}\Big\vert_{\mbox{\scriptsize oscl}^{(1)}\equiv 0}
+F_{p^2}\right)(2\pi)^{d+1}
\delta\Bigl(p_1+\sum_{s=2,3}p_s\Bigr)
\delta\Bigl(\e+\sum_{s=2,3}\alpha_s\Bigr)
\nn\\
&&=\abs{\RT{2}}^2\Biggl\{1+\frac{\e}{\alpha_2}\left(
1-\wtNFP^{(2)}\right)\Biggr\}
\nn\\
&&\qquad\times
:\exp\left\{-ip_1^\mu\sqrt{\pi}X^{(2)\mu}(\sigma=0)
\right\}:\left(1 + \e\Pdrv{}{\alpha_2}\right)
\ket{r(2,3)} ,
\label{eq:mu-F+F-dp-da}
\end{eqnarray}
where $\wtNFP$ is given by (\ref{eq:wtNFP}), and $\ket{r(2,3)}$ is
the reflector (without the $\ac_0$-part),
\begin{eqnarray}
&&\ket{r(2,3)}\equiv\exp\left\{
-\sum_\pm\sum_{n\ge 1}(-)^n\left(
\frac{1}{n}\a_{-n}^{(\pm)(2)}\cdot\a_{-n}^{(\pm)(3)}
- c_{-n}^{(\pm)(2)}\ac_{-n}^{(\pm)(3)}
+ \ac_{-n}^{(\pm)(2)}c_{-n}^{(\pm)(3)}\right)\right\}
\ket{0}_{2,3}
\nn\\
&&\qquad\qquad\times (2\pi)^{d}\delta^{d}\left(p_2+p_3\right)
\cdot 2\pi\delta\left(\alpha_2+\alpha_3\right) ,
\label{eq:r}
\end{eqnarray}
which enjoys the following property,
\begin{equation}
\pmatrix{\ds \a_n^{(\pm)(2)}+(-)^n\a_{-n}^{(\pm)(3)}\cr
\ds c_n^{(\pm)(2)}+(-)^n c_{-n}^{(\pm)(3)} \cr
\ds \ac_n^{(\pm)(2)}-(-)^n\ac_{-n}^{(\pm)(3)}}\ket{r(2,3)}=0
\label{eq:(a+a)r=0}
\quad(n=\pm 1, \pm 2, \ldots) .
\end{equation}
Two comments are in order for eq.\ (\ref{eq:mu-F+F-dp-da}).
First, as seen from eq.\ (\ref{eq:F_p^2-exp}), $F_{p^2}$ does not
contribute to (\ref{eq:ToBeCalc}) and hence is omitted on the RHS
of (\ref{eq:mu-F+F-dp-da}). However, when we consider $\zeta^\mu(x)$
and $\xi^\mu(x)$ of higher power than linear in $x^\mu$, we have to
take $F_{p^2}$ into account.
Second, $\sqrt{\pi}X^{(2)\mu}(\sigma=0)$ in (\ref{eq:mu-F+F-dp-da})
originally appears as
\begin{equation}
i\left(\pdrv{}{p_2^\mu}\right)
- (i/2)\sum_\pm\sum_{n\ge 1}\frac{1}{n}\left(
\a_{-n}^{(\pm)\mu(2)}+(-)^n\a_{-n}^{(\pm)\mu(3)}\right) ,
\label{eq:X_creation}
\end{equation}
consisting solely of the creation operators. Therefore, we need the
normal ordering symbol in eq.\ (\ref{eq:mu-F+F-dp-da}).

\noindent
\underline{Formulas for contractions}

Our next task is to obtain the expressions of various ``contractions''
which appear in the calculation of
${}_1\kern-3pt\bra{0}\cdots\ket{0}_{1,2,3}$ in (\ref{eq:A}).
First, we have
\begin{eqnarray}
&&\wick{2}{<1\a_1^{(\pm)\mu(1)}\ >1F}=
\sum_{s=2,3}\sum_{m\ge 1}\N{1m}{1s}\a_{-m}^{(\pm)\mu(s)}
+ \Half\N{1}{1}\P^\mu
\nn\\
&&\phantom{\wick{2}{<1\a_1^{(\pm)\mu(1)}\ >1F}}
=\abs{\IRT{2}}\left\{
\sum_{n\ge 1}\left(\a_{-n}^{(\pm)\mu(2)} -(-)^n\a_{-n}^{(\pm)\mu(3)}
\right) +\Half p_2^\mu
-\left(\frac{\alpha_2}{2\e}-\frac{1}{4}\right)p_1^\mu
+ O(\e)\right\} ,
\nn\\
&&\label{eq:wick-a-F}\\
&&\wick{1}{<1{\ac}_1^{(\mp)(1)}\ >1F}=
\sum_{s=2,3}\sum_{m\ge 1}\N{1m}{1s}\frac{\e}{\alpha_s}
\ac_{-m}^{(\mp)(s)}\nn\\
&&\phantom{\wick{1}{<1{\ac}_1^{(\mp)(1)}\ >1F}}
=\sum_{m\ge 1}\abs{\IRT{2}}\frac{\e}{\alpha_2}\left(
\ac_{-m}^{(\mp)(2)} + (-)^m\ac_{-m}^{(\mp)(3)}\right)
+O(\e^3) .
\label{eq:wick-ac-F}
\end{eqnarray}
Using the expansion (c.f., eq.\ (A.6) of \cite{HataNagoshi})
\begin{equation}
w_{1,1}^{(r)}=
(-)^{r-1}\frac{1}{\e}\abs{\IRT{2}}\left(
1 - \frac{\e}{2\alpha_{2}} + O(\e^2)\right) ,
\end{equation}
for the coefficient $w_{1,1}^{(r)}$ ($r=2,3$) in
\begin{equation}
w_I^{(r)}\equiv\frac{i}{\sqrt{2}}\sum_\pm\sum_{s=1}^3\sum_{n\ge 1}
w_{n,s}^{(r)}\gamma_{-n}^{(\pm)(s)} ,
\label{eq:w_I^r}
\end{equation}
we get
\begin{eqnarray}
&&\wick{2}{<1{\ac}_1^{(\mp)(1)} >1w_I^{(r)}}
=\wick{2}{<1{\ac}_1^{(\mp)(1)} \frac{i}{\sqrt{2}}w_{1,1}^{(r)}
>1\gamma_{-1}^{(\mp)(1)}}
=\frac{1}{\sqrt{2}}\,\e w_{1,1}^{(r)}\nn\\
&&\phantom{\wick{2}{<1{\ac}_1^{(\mp)(1)} >1w_I^{(r)}}}
=(-)^{r-1}\frac{1}{\sqrt{2}}\abs{\IRT{2}}\left(
1 - \frac{\e}{2\alpha_{2}} + O(\e^2)\right)
\quad (r=2,3) .
\end{eqnarray}
{}From
\begin{equation}
w_{n,s}^{(r)}=(-)^r\frac{1}{n\alpha_2}\times
\cases{1 &($s=2$) \cr (-)^n &($s=3$)} +O(\e)
\quad (r=2,3) ,
\label{eq:w_ns^r}
\end{equation}
which is a special case of eq.\ (A.6) of \cite{HataNagoshi},
we have for $w_I^{(r=2,3)}$ left after the contractions:
\begin{eqnarray}
&&\wt{w}_I^{(r=2,3)}\equiv w_I^{(r)}\Big\vert_{c^{(1)}=0}
=\frac{i}{\sqrt{2}}\sum_\pm\sum_{s=2,3}\sum_{n\ge 1}
w_{n,s}^{(r)}\gamma_{-n}^{(\pm)(s)}
\nn\\
&&\qquad\qquad
=(-)^r\isqrt{2}\sum_\pm\sum_{n\ge 1}
\left(c_{-n}^{(\pm)(2)}-(-)^n c_{-n}^{(\pm)(3)}
\right) + O(\e) .
\label{eq:wt_w}
\end{eqnarray}

\noindent
\underline{Contractions I : $\wick{1}{<1\a_1^{(\pm)\mu(1)} >1 F}\cdot
\wick{1}{<1{\ac}_1^{(\mp)(1)} >1 w_I^{(r=2,3)}}$}

Let us consider all possible contractions which appear in
${}_1\kern-3pt\bra{0}\cdots\ket{0}_{1,2,3}$ of (\ref{eq:A}).
First is the contraction of the type
$\wick{1}{<1\a_1^{(\pm)\mu(1)} >1 F}\cdot
\wick{1}{<1{\ac}_1^{(\mp)(1)} >1 w_I^{(r=2,3)}}$.
It has contributions from the second and the last terms on the RHS of
\begin{equation}
\prod_{r=2,3}\left(1 - \ac_0^{(r)}\frac{1}{\sqrt{2}}w_I^{(r)}\right)
=1+\sum_{r=2,3}\left(-\ac_0^{(r)}\isqrt{2}w_I^{(r)}\right)
+\ac_0^{(2)}\isqrt{2}w_I^{(2)}\ac_0^{(3)}\isqrt{2}w_I^{(3)} .
\label{eq:prod(1-acw)}
\end{equation}
The contribution from the second term is
\begin{eqnarray}
&&\wick{43}{<1\a_1^{(\pm)\mu(1)}<2{\ac}^{(\mp)(1)}
\sum_{r=2,3}\Bigl(-\ac_0^{(r)}\isqrt{2}>2w_I^{(r)}\Bigr)>1F}
=-\left(\ac_0^{(2)}-\ac_0^{(3)}\right)
\Half\left(\IRT{2}\right)^2\left(1-\frac{\e}{2\alpha_2}\right)
\nn\\
&&\qquad\qquad\times\left\{
\sum_{n\ge 1}\left[
\a_{-n}^{(\pm)\mu(2)}-(-)^n\a_{-n}^{(\pm)\mu(3)}\right]
+\Half{}p_2^\mu-\left(\frac{\alpha_2}{2\e}-\frac{1}{4}\right)p_1^\mu
\right\} .
\label{eq:aFacw}
\end{eqnarray}
Corresponding to $\bra{\Lambda_-}$ ($\bra{\Lambda_+}$), we
anti-symmetrize (symmetrize) (\ref{eq:aFacw}) with respect to $(\pm)$:
\begin{eqnarray}
&&\mbox{Anti-symm.: }
\sum_\pm(\pm)\mbox{(\ref{eq:aFacw})}
=\sqrt{\pi}\left(\ac_0^{(2)}-\ac_0^{(3)}\right)
\left(\IRT{2}\right)^2
\left.\Drv{}{\sigma}X^{\mu(2)}(\sigma)\right|_{\sigma=0} ,
\label{eq:aFacw_antisym}\\[5pt]
&&\mbox{Symm.: }\sum_\pm\mbox{(\ref{eq:aFacw})}=
- \Half\left(\ac_0^{(2)}-\ac_0^{(3)}\right)
\left(\IRT{2}\right)^2\left(1-\frac{\e}{2\alpha_2}\right)
\nn\\
&&\hspace*{5cm}\times
\left\{
2\sqrt{\pi}P_\mu^{(2)}(\sigma=0)
-\left(\frac{\alpha_2}{\e}-\Half\right)p_1^\mu
\right\} .
\label{eq:aFacw_sym}
\end{eqnarray}
In eqs.\ (\ref{eq:aFacw_antisym}) and (\ref{eq:aFacw_sym}), we have
converted the creation operators of the string 3 into
the annihilation ones of the string 2 using (\ref{eq:(a+a)r=0})
and hence we have $(d/d\sigma)X^{\mu(2)}$
and $P_\mu^{(2)}$ there.

Next, as the contribution from the last term of (\ref{eq:prod(1-acw)}),
we obtain
\begin{eqnarray}
&&\wick{2}{<1\a_1^{(\pm)\mu(1)} >1 F}\left(
\wick{3}{<1{\ac}^{(\mp)(1)}\isqrt{2}\ac_0^{(2)}>1 w_I^{(2)}
\isqrt{2}\ac_0^{(3)}\wt{w}_I^{(3)}}
+\wick{3}{<1{\ac}^{(\mp)(1)}\isqrt{2}\ac_0^{(2)}\wt{w}_I^{(2)}
\isqrt{2}\ac_0^{(3)}>1 w_I^{(3)}}\right)
\nn\\
&&=-\Half\wick{2}{<1\a_1^{(\pm)\mu(1)} >1 F}\,
\wick{2}{<1{\ac}^{(\pm)(1)} >1 w_I^{(2)}}\left(
\wt{w}_I^{(3)} + \wt{w}_I^{(2)}\right)\ac_0^{(2)}\ac_0^{(3)}+
\ldots
\nn\\
&&=O(\e^2)\times p_1^\mu\times
\left(\mbox{terms linear in }c_{-n}^{(2,3)}\right) ,
\label{eq:aFacw_acw}
\end{eqnarray}
where we have used the fact that
$\wt{w}_I^{(3)} + \wt{w}_I^{(2)}=O(\e)$
as seen from (\ref{eq:wt_w}). Eq.\ (\ref{eq:aFacw_acw})
is manifestly symmetric with respect to $(\pm)$.

\noindent
\underline{Contractions II :
$\wick{1}{<1\a_1^{(\pm)\mu(1)} >1 F}\cdot
\wick{1}{<1{\ac}_1^{(\mp)(1)} >1 F}$}

Since $\wt{w}_I^{(2)}\wt{w}_I^{(3)}=O(\e)$, we have
\begin{eqnarray}
&&\prod_{r=2,3}\left(1-\ac_0^{(r)}\isqrt{2}\wt{w}_I^{(r)}\right)
= 1 +
\sum_{r=2,3}\Bigl(-\ac_0^{(r)}\isqrt{2}\wt{w}_I^{(r)}\Bigr)
-\Half\ac_0^{(2)}\ac_0^{(3)}\wt{w}_I^{(2)}\wt{w}_I^{(3)}
\nn\\
&&=1 -\isqrt{2}\left(\ac_0^{(2)}-\ac_0^{(3)}\right)\wt{w}_I^{(2)}
+O(\e) ,
\end{eqnarray}
and hence
\begin{eqnarray}
&&\wick{43}{<1\a_1^{(\pm)\mu(1)}<2{\ac}^{(\mp)(1)}
\prod_{r=2,3}\left(1-\ac_0^{(r)}\isqrt{2}\wt{w}_I^{(r)}\right)
>1 F >2 F} \nn\\
&&=O(\e^2)\times p_1^\mu\times \left(\mbox{terms linear in }
c_{-n}^{(2,3)}\right)\nn\\
&&
-\frac{1}{4}p_1^\mu\abs{\IRT{2}}^2\left(\ac_0^{(2)}-\ac_0^{(3)}\right)
\sum_{m\ge 1}\left(
\ac_{-m}^{(\mp)(2)} + (-)^m\ac_{-m}^{(\mp)(3)}\right)
\sum_\pm\sum_{n\ge 1}\left(
c_{-n}^{(\pm)(2)}-(-)^n c_{-n}^{(\pm)(3)}\right) .
\nn\\
\label{eq:aFacF}
\end{eqnarray}
Among many terms on the RHS of (\ref{eq:wick-a-F}), only the
$O(\e^0)$ term,
$\abs{\e/e\alpha_2}\left(-\alpha_2/2\e\right)p_1^\mu$,
contributes to (\ref{eq:aFacF}).
(Anti-)symmetrizing (\ref{eq:aFacF}) with respect to $(\pm)$, we
obtain
\begin{eqnarray}
&&\mbox{Anti-symm.: }
\sum_\pm(\pm)\mbox{(\ref{eq:aFacF})}
=O(\e^2)\times p_1^\mu\times \left(
\mbox{terms linear in } c_{-n}^{(2,3)}\right)\nn\\
&&\qquad
-\pi{}p_1^\mu\abs{\IRT{2}}^2\left(\ac_0^{(2)}-\ac_0^{(3)}\right)
:i\pi_c^{(2)}|_{\rm oscl}(\sigma=0)\cdot
i\pi_{\ac}^{(2)}|_{\rm oscl}(\sigma=0):\ ,
\label{eq:aFacF_antisym}\\
&&\mbox{Symm.: }
\sum_\pm\mbox{(\ref{eq:aFacF})}
=O(\e^2)\times p_1^\mu\times \left(
\mbox{terms linear in } c_{-n}^{(2,3)}\right)\nn\\
&&\qquad
-\pi{}p_1^\mu\abs{\IRT{2}}^2\left(\ac_0^{(2)}-\ac_0^{(3)}\right)
:\ac|_{\rm oscl}(\sigma=0)\cdot
i\pi_{\ac}|_{\rm oscl}(\sigma=0): \ ,
\label{eq:aFacF_sym}
\end{eqnarray}
where ``oscl'' indicates the creation/annihilation operator parts.

\noindent
\underline{Evaluation of (\ref{eq:ToBeCalc})}

Having obtained all the necessary formulas, let us evaluate
(\ref{eq:ToBeCalc}). In the following we use the abbreviation
$\wp^{(2,3)}\equiv \wp^{(2)}\wp^{(3)}$.
First is (\ref{eq:ToBeCalc}) for $\Lambda_-$:
\begin{eqnarray}
&&\int d1 \braket{\Lambda_-(1)}{V(1,2,3)}
=a_{\mu\nu}\wp^{(2,3)}
\lim_{\e\to 0}\lim_{p_1\to 0}\Pdrv{}{p_1^\nu}\Bigl(
\mbox{(\ref{eq:aFacw_antisym})}
+ \mbox{(\ref{eq:aFacF_antisym})}
\Bigr)\times
\mbox{(\ref{eq:mu-F+F-dp-da})}\nn\\
&&=a_{\mu\nu}\wp^{(2,3)}\left(-i\pi:\!\Drv{X^{\mu}}{\sigma}
\cdot X^{\nu}\!: - \pi\eta^{\mu\nu}
:\!i\pi_c|_{\rm oscl}\,i\pi_{\ac}|_{\rm oscl}\!:
\right)^{\!(2)}\Bigg\vert_{\sigma=0}
\ket{R(2,3)}\nn\\
&&=\Half\,a_{\mu\nu}\left(-i\int_0^{2\pi}\!\!d\sigma\,
\Drv{X^{\mu}(\sigma)}{\sigma}X^{\nu}(\sigma)
-\eta^{\mu\nu}\int_0^{2\pi}\!\!d\sigma\,
i\pi_c(\sigma)|_{\rm oscl}\cdot{}i\pi_{\ac}(\sigma)|_{\rm oscl}
\right)^{\!(2)}
\ket{R(2,3)} ,
\nn\\
\label{eq:ans_antisym}
\end{eqnarray}
where $\ket{R(2,3)}$ is the full reflector,
\begin{equation}
\ket{R(2,3)}\equiv
\ket{r(2,3)}\times\left(\ac_0^{(2)}-\ac_0^{(3)}\right) .
\label{eq:R}
\end{equation}
In obtaining (\ref{eq:ans_antisym}) we have used the fact that
the first term on the RHS of (\ref{eq:aFacF_antisym}) does not
contribute due to the presence of $\wp^{(2,3)}$.

Next, (\ref{eq:ToBeCalc}) for $\Lambda_+$ is
\begin{eqnarray}
&&\int\! d1 \braket{\Lambda_+(1)}{V(1,2,3)}
=b_{\mu\nu}
\wp^{(2,3)}\lim_{\e\to 0}\lim_{p_1\to 0}\Pdrv{}{p_1^\nu}
\Bigl(
\mbox{(\ref{eq:aFacw_sym})}
+\mbox{(\ref{eq:aFacw_acw})}
+ \mbox{(\ref{eq:aFacF_sym})}\Bigr)\times
\mbox{(\ref{eq:mu-F+F-dp-da})}\nn\\
&&=b_{\mu\nu}
\wp^{(2,3)}\left\{i\pi:P_\mu^{(2)}X_\nu^{(2)}:
\!\! - \pi\eta^{\mu\nu}:\!
\ac^{(2)}|_{\rm oscl}\,i\pi_{\ac}^{(2)}|_{\rm oscl}\!:
+\Half\eta^{\mu\nu}\left(\alpha_2\Pdrv{}{\alpha_2}
-\wtNFP^{(2)}\right)\right\}_{\sigma=0}\ket{R(2,3)}
\nn\\
&&=\Half{}b_{\mu\nu}\left(
\frac{i}{2}\int_0^{2\pi}\!\!d\sigma
\Bigl\{P_\mu(\sigma),X_\nu(\sigma)\Bigr\}
+\Half\eta^{\mu\nu}\left\{\alpha, \Pdrv{}{\alpha}\right\}
-\eta_{\mu\nu}\G
\right)^{\!(2)}\ket{R(2,3)} .
\label{eq:ans_sym}
\end{eqnarray}
The points in deriving (\ref{eq:ans_sym}) are as follows:
\begin{itemize}
\item Due to the $-(\alpha_2/\e)p_1^\mu$ term in
  (\ref{eq:aFacw_sym}), we have to take into account the
  next-to-leading terms in (\ref{eq:mu-F+F-dp-da}) and
  (\ref{eq:aFacw_sym}). In particular,
  $\e\alpha_2\left(\pdrv{}{\alpha_2}\right)$ in
  (\ref{eq:mu-F+F-dp-da}) does contribute.

\item $\delta(\alpha)$ in $\bra{\Lambda_\pm}$ (\ref{eq:Lambda}) should
  be understood to imply
\begin{equation}
\delta(\alpha)\equiv \lim_{\e\to +0}\Half\Bigl(
\delta(\alpha-\e)+\delta(\alpha+\e)\Bigr) ,
\label{eq:delta_alpha}
\end{equation}
owing to the hermiticity of $\Lambda_\pm$.
Therefore, the $O(1/\e)$ term originating from the term
$-(\alpha_2/\e)p_1^\mu$ in (\ref{eq:aFacw_sym}) is missing from
(\ref{eq:ans_sym}).

\item Neither the first term on the RHS of (\ref{eq:aFacF_sym}) nor
(\ref{eq:aFacw_acw}) contribute owing to the presence of
$\wp^{(2,3)}$.
\end{itemize}

Eqs.\ (\ref{eq:Psi*L-}) and (\ref{eq:Psi*L+}) are consequences of
eqs. (\ref{eq:ans_antisym}) and (\ref{eq:ans_sym}) as well as
the formula for the star product,
\begin{equation}
\ket{\left(\Psi\star\Lambda\right)(2)}
=\epsilon_{\Psi}\epsilon_{\Lambda}\int\!d3\!\int\!d1
\bra{\Psi(3)}\bra{\Lambda(1)}\!\ket{V(1,2,3)} ,
\label{eq:Psi*Lambda}
\end{equation}
where the sign factor $\epsilon_\Psi$ is $1$ ($-1$) if $\Psi$ is
hermitian (anti-hermitian):
\begin{equation}
\bra{\Psi(2)}=\epsilon_\Psi\int\!d1\braket{R(1,2)}{\Psi(1)} .
\label{eq:hermiticity}
\end{equation}
The transformation functional $\Lambda$ is anti-hermitian and we have
$\epsilon_\Lambda=-1$. Note that the integration measure $\int\!d1$ is
Grassmann-odd and anti-hermitian,
$\left(\int\!d1\right)^\dagger=-\int\!d1$, and hence eq.\
(\ref{eq:hermiticity}) implies
$\ket{\Psi(2)}=\epsilon_\Psi(-)^{\abs{\Psi}}
\int\!d1\braket{\Psi(1)}{R(1,2)}$, where $\abs{\Psi}=0$ ($1$) if
$\Psi$ is Grassmann-even (-odd).

\newpage
%%%%%%%%%% References %%%%%%%%%%%%%%%%%%%%%%%%%
\newcommand{\J}[4]{{\sl #1} {\bf #2} (#3) #4}
\newcommand{\andJ}[3]{{\bf #1} (#2) #3}
\newcommand{\AP}{Ann.\ Phys.\ (N.Y.)}
\newcommand{\MPL}{Mod.\ Phys.\ Lett.}
\newcommand{\NP}{Nucl.\ Phys.}
\newcommand{\PL}{Phys.\ Lett.}
\newcommand{\PR}{Phys.\ Rev.}
\newcommand{\PRL}{Phys.\ Rev.\ Lett.}
\newcommand{\PTP}{Prog.\ Theor.\ Phys.}
\newcommand{\HIKKO}{
H.\ Hata, K.\ Itoh, T.\ Kugo, H.\ Kunitomo and K.\ Ogawa}
%%%%%%%%%%%%%%%%%%%%%%%%%%%%%%%%%%%%%%%%%%%%%%%

\end{document}